\documentclass{article}
\usepackage{epsfig,amssymb}
\usepackage[dvips]{color}

\begin{document}

\newtheorem{theor}{Theorem} \newtheorem{defin}[theor]{Definition}
\newtheorem{corol}[theor]{Corollary} \newtheorem{prop}{Proposition}
\newtheorem{lem}[theor]{Lemma} \newcommand{\blem}{\begin{lem}}
\newcommand{\elem}{\end{lem}} \newcommand{\bp}{\begin{prop}}
\newcommand{\ep}{\end{prop}} \newtheorem{example}{Example}
\newcommand{\bex}{\begin{example}} \newcommand{\eex}{\end{example}}
\newcommand{\sq}{\lhd\!\!\!\rhd} \newcommand{\bt}{\begin{theor}}
\newcommand{\et}{\end{theor}}   \newcommand{\bd}{\begin{defin}}
\newcommand{\ed}{\end{defin}}  \newcommand{\bco}{\begin{corol}}
\newcommand{\eco}{\end{corol}}   \newcommand {\6}{\\[.6em]}  \newcommand
{\2}{\\[-2em]}  \newcommand {\be}{\begin{enumerate}}  \newcommand
{\ee}{\end{enumerate}}        
 \newcommand {\bi}{\begin{itemize}}  \newcommand{\ei}{\end{itemize}}   
 \newcommand{\I}{\item}

\title{\bf 
Fuzzy Relational Modeling of Cost and Affordability for Advanced Technology 
Manufacturing Environment}
\author{{\normalsize Ladislav J. Kohout\footnote{All correspondence to kohout@cs.fsu.edu} \ and \ Eunjin Kim\footnote{Current Affiliation:Prof. Eunjin Kim, Dept. of Computer Science, John D. Odergard School of Aerospace Sciences, University of North Dakota, Grand Forks, ND 582002--9015, USA. }} \\
{\normalsize Dept. of Computer Science, Florida State University, Tallahassee, FL 32306-4530} \\
{\normalsize  kohout@cs.fsu.edu, \ \ ejkim@cs.fsu.edu } \\
{\normalsize Gary  Zenz} \\
{\normalsize College of Business, Florida State University, Tallahassee, FL 32306-1110} \\ 
{\normalsize gzenz@cob.fsu.edu }}
   
\date{}

\maketitle

\begin{quotation}
\textit{This paper originally appeared in 1999 NSF
  Design \& Manufacturing Grantees \\  
Conference Proceedings. The table of contents and refrences to
  publications \\
originally referred to as ``in press'' have been added when available. }
\end{quotation}

\begin{abstract}
Relational representation of knowledge
 makes it possible to perform all the computations and 
decision making in a uniform relational way \cite{92.7}, by means 
of {\it special relational compositions} called triangle and square 
products. These were first introduced by Bandler and Kohout in 1977 
\cite{87.8},\cite{80.3},\cite{77.rel}
and are referred to as the BK-products  in the literature 
\cite{hajek.bk},\cite{deb+ker93pr},\cite{deb+kerMP}. Their 
theory and applications have made substantial progress since then.  

 BK-relational product can be used to compare relational 
 structures. 
  Relations so constructed might exhibit some important
relational properties that reveal important characteristics
and interrelationships of the source of information from which
they were generated. Hence, methods for detecting various relational 
properties of given relations are important.

Collecting engineering data concerning various manufacturing processes,
parts, subsystems and manufactured goods is usually done by
physical measurements  
of such physical entities that serve as {\it cost drivers}. 
Because one of major  concerns is to deal with affordability
issues also in the situations when such ``hard" data are not available, 
relational analysis on data and knowledge can be elicited by questioning
engineers. A case study of this kind is described in Sec. 3.  Here,
instead of physical measurement devices  we use psychometric tools
invented by behavioral scientists called {\it repertory grids}
(RPG). Our relational analysis can be used to analyze data
(e.g. process parameters) collected by physical measurements as well
as data obtained by knowledge elicitation from human experts. 

Relational properties characterizing the structure of knowledge, such as 
 reflexivity,  symmetry, and transitivity, and classes such as tolerances, 
 equivalences and partial orders can be extracted from the linguistic 
 information elicited by repertory grids. 

Testing the fuzzy relational structure for various relational properties
allows us to discover dependencies, hierarchies, 
similarities, and equivalences of
the attributes characterizing technological processes and manufactured
artifacts  in their relationship to costs and performance. 

How to use our methods for ranking of various technologies with
respect to affordability is shown in Sec. 4. In section 5, a more
detailed study of cost drivers by means of fuzzy relational products
is described. 

A brief overview of mathematical aspects of BK-relational products is
given in Appendix 1 together with further references in the
literature.

\end{abstract}

\pagebreak 
\noindent 
\textbf{Acknowledgement of Support and Disclaimer}\\
\textit{"This material is based upon work supported by the National Science
Foundation under Grant No. DMI 952 5991."\\[-0.9em]}

\textit{``Any opinions, findings, and conclusions or recommendations expressed in this material are those of the author(s) and do not necessarily reflect the views of the National Science Foundation.''}

{\small \tableofcontents}

\section{Importance of the Assessment of Cost and Affordability}

\subsection{Background, Goals, and Methods Used}

In an advanced 
technology environment, the {\it key to achieving affordability goals} 
which are necessary
to maintain a competitive position of the US industry in the domestic and world markets,
is to deal with complicating uncertainties in materials, 
fabrication, and manufacturing. 

Pratt\&Whitney, our industrial partner cooperating with
us in this project, is one of the companies that belongs to the United 
Technologies group, a diversified producer of consumer, commercial, and
military products. Pratt\&Whitney is one of the firms  that are at the 
leading edge of high technology \cite{97.1}. 
Such companies face a formidable problem, being often forced to make 
technological and business decisions based 
on {\it incomplete, uncertain information} about the product that is yet 
{\it to be} designed and manufactured. The industry  needs affordability 
models applicable to such manufacturing problems.  Scarcity of information 
concerning untried technologies and the lack of historical data base 
{\it are  the main characteristics} of this problem. 

Our current program addresses this objective.  Working jointly with 
our industrial partner Pratt \& Whitney, we have
developed practical fuzzy relational techniques that can assist in
affordability modeling  interfaced with engineering design
methods.  Particularly important is identifying dependencies, hierarchies, 
similarities and equivalences of attributes characterizing processes and 
products in their relationship to cost and performances. 

The importance of our techniques stems from the fact that they have been
developed for the situations where often only incomplete information and
small data sets are available.  This is important for strategies for
integration of cost into design at very early stages, and for advanced
technological design of products never before manufactured.

\subsection{The Long Term Objectives of Our Work}

Jointly with our industrial partner Pratt\&Whitney we have formulated  
the objectives for our {\bf long term cooperation} that are listed below. 
Namely, we attempt to: \\

\noindent 
{\bf (LTO-1)} Provide a systematic framework for integrating engineering 
design and manufacturing activities with the management, organizational, 
accounting and financial activities of the enterprise. \\
{\bf (LTO-2)} Identify all technical, human and organizational 
contribution to costs.\\
{\bf (LTO-3)} Deal systematically with incomplete, uncertain 
and conflicting information, constraints and consequences. \\
{\bf (LTO-4)} Deal with uncertainty in estimates, and incorporate the 
estimates concurrently into engineering design. \\
{\bf (LTO-5)} All the methodologies and techniques resulting from the objectives 
LTO-1 to LTO-4 have to represent data and knowledge in the form compatible 
with the framework needed for design of computer information systems -- 
preferably computer based distributed Intelligent Systems for 
manufacturing and telemanufacturing. \\
{\bf (LTO-6)} The techniques and methodologies should be compatible with 
a high level conceptual model of cooperating industrial firm, reflecting 
the the effect of interaction of cooperating firms on the cost  and
affordability of products. This should include the effect of procurement,
purchasing and marketing. \\[-0.2 cm]

To deal with LTO-1 and LTO-4 we use Fuzzy relational mathematics. 
This  provides 
a framework for working with incomplete and/or conflicting information, 
constraints, consequences; and also with uncertainty of probabilistic  
as well as of non-probabilistic nature 
\cite{97.9},\cite{97.4},\cite{96.7},\cite{96.10}.

To integrate human factors with technological ones (objective LTO-2) 
one has to take into account not only the technological 
design and production concepts and data, but also the psychological 
and linguistic constructs utilized by human participants. This requires 
special techniques we have developed 
\cite{92.11},\cite{96.10},\cite{89.21},\cite{89.20},\cite{89.2},\cite{86.19}. 
Value analysis method \cite{zenz.bk},[R12],[R15] has provided the bridge 
for incorporating management, financial and organizational activities 
(objective LTO-1).

Uniform data and knowledge representation equally applicable to  
objectives LTO-1, LTO-2, LTO-5 has been provided by the methodology 
of Activity Structures \cite{89.4},\cite{89.5} which includes relational 
data and knowledge representation as its integral part. 

The much required unification of data analysis and computational methods 
we have achieved by combining relational mathematics 
\cite{87.8},\cite{86.2} and computational science
\cite{96.3},\cite{97.8},\cite{96.13},\cite{96.14} within the framework
of relational virtual computer architectures  
\cite{85.3},\cite{93.1},\cite{89.23},\cite{96.12}. 
In particular, fuzzy relations, 
BK-relational products\footnote{BK-products is a term used in the literature 
on fuzzy sets to designate new relational compositions discovered by Bandler 
and Kohout in 1977.}  
\cite{87.8},\cite{97.5},\cite{96.2},\cite{91.15},\cite{95.13},\cite{94.5} 
and fast fuzzy relational algorithms 
\cite{88.3},\cite{92.7},\cite{92.14} have been consistently used for 
data analysis, knowledge elicitation, knowledge and data representation and 
further information processing. For recent results see
publications\footnote{
References starting with R appear in the list of 
publications originated form this grant listed in Appendix 2.} [R4], [R8],
[R10] resulting from the grant support (NSF DMI 952 5991)  listed in
Appendix 2 .

Finally, integration of all the information and knowledge dealt with 
in our objectives into a global system that 
synthesize the information 
relevant to affordability analysis is based on Activity Structures 
methodology \cite{93.12},\cite{93.12},\cite{94.3},\cite{96.12}, 
This methodology was created to give a unified platform 
for development of distributed intelligent systems \cite{89.5}, hence  
it has been used to achieve the objectives LTO-5 and LTO-6. 
For recent results see publications 
[R10] and  [R12] resulting from the  grant support  (NSF DMI 9525991) listed
in Appendix 2.

\subsection{The Summary our work supported by the current
grant  NSF DMI 9525991} 

Within the framework of the long term objectives outlined in the previous
section we have worked out more detailed  objectives for our current
projects.  
In this section we discuss the specific objectives our current NSF grant
(NSF DMI 9525991) entitled 
 {\it Decision-Making with Incomplete Information in an Integrated 
Product and Process Development Enterprise -- A Management 
Decision Tool for Cost Modeling and Affordability 
Applications}. Our industrial partner for this work has been Pratt \&
Whitney. 

The main objectives for the 3 years of the current grant for the period 
October 1995 -- September 1998 are as follows: 

\begin{enumerate}
\item Use of Fuzzy Relational Methods for data and knowledge elicitation 
and representation, and affordability modeling. 
\item Value Analysis for Integration of Technology and Business.
\item Problems of Engineering Design: prototype software system for 
estimating product/process
cost based on the fuzzy multi-attribute utility theory.
\item Comparison of Fuzzy and Probabilistic Methods in their 
applicability to affordability data.
\item Knowledge Transfer to Industry and Education. 
\end{enumerate}

As we are concerned in this paper with fuzzy relational knowledge
representation techniques and data analysis with imprecise and incomplete
data we discuss in the sequel objectives (1) and (2) in detail. For
information on other objectives the reader is referred to the following 
papers:  \cite{97.2},\cite{97.1},\cite{97.12}.

\section{FRASMod  Relational Affordability Knowledge 
Representation Structure}

The contexts addressed in this project that are of particular interest to
our industrial partner Pratt\&Whitney \cite{97.1} are depicted in Figure 1.

\begin{figure}
\includegraphics[width=6in]{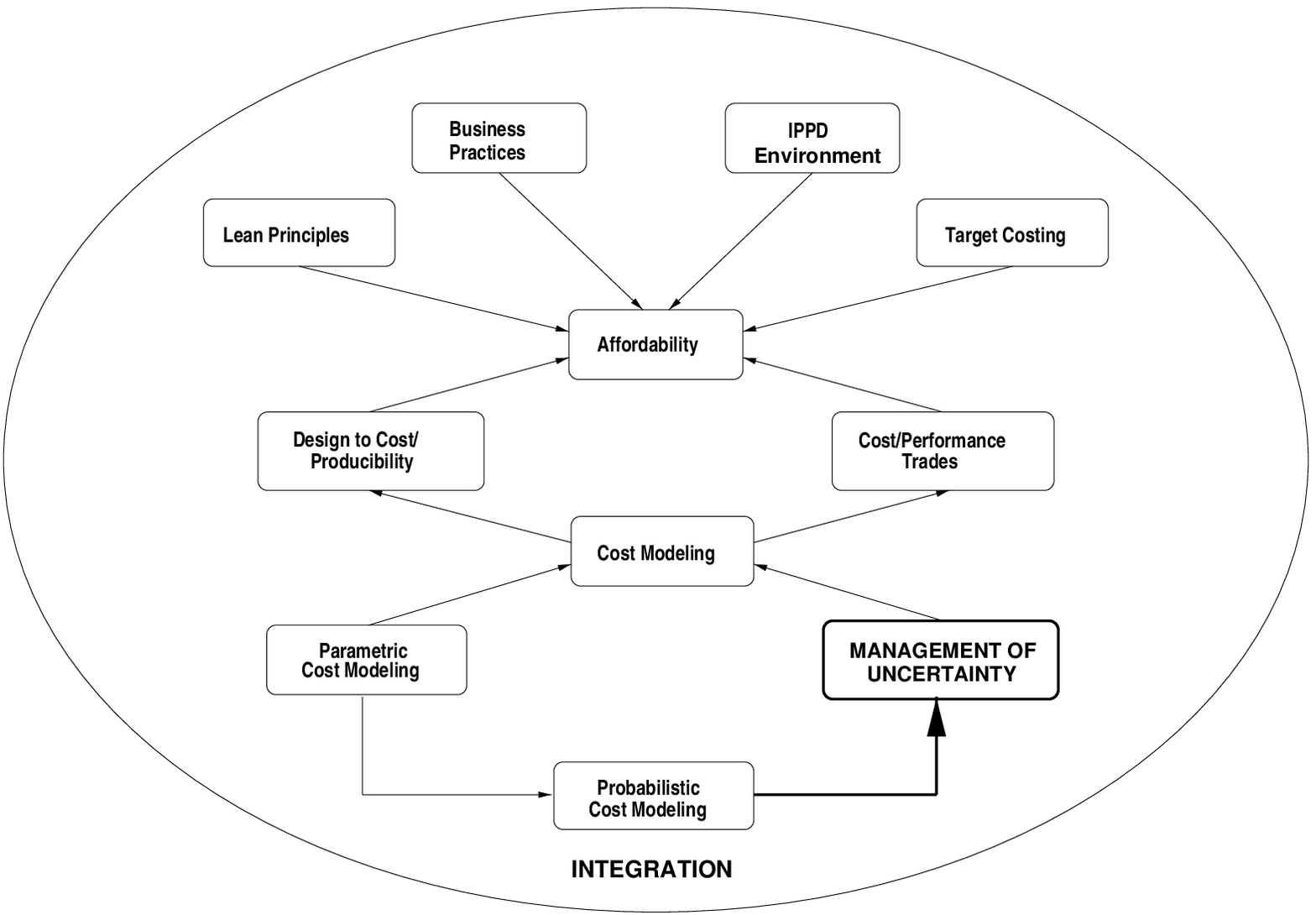}
\caption{}
\end{figure}

Out of 11 subsystems forming the Industrial context of Affordability 
modeling shown in this figure  we identified six pivotal issues that 
have been addressed while developing our relational affordability 
representation scheme.  These are as follows:           
                                                         
\noindent
\hspace*{ 0.9 cm} \(\bullet\) Affordability; \\
\hspace*{ 0.9 cm} \(\bullet\) Management of Uncertainty; \\
\hspace*{ 0.9 cm} \(\bullet\) Cost interval and fuzzy modeling; \\
\hspace*{ 0.9 cm} \(\bullet\) Cost/Performance Trades, \\
\hspace*{ 0.9 cm} \(\bullet\) IPPD Environment activities \\
\hspace*{ 0.9 cm} \(\bullet\) Business Practices.

The Fuzzy Relational Affordability Systemic Model (FRASMod)  we have 
developed is designed to capture and integrate the above listed 6 
perspectives of manufacturing activities within a unified knowledge
representation structure. 
 
In the use of FRASMod the key entities of each perspective are 
identified using the exploratory knowledge elicitation  and mapped into 
a relational subsystem and a relational coupling structure that shows 
potential interactions of the entities corresponding to different 
perspectives.

We have included the following conceptual categories (i.e. semiotic 
descriptors \cite{89.5},\cite{89.4},\cite{96.10}) of relations 
in the knowledge representation structure \cite{97.11} used in FRASMod: 

\noindent 
\hspace*{ 0.9 cm} \(\bullet\) Objects, \\
\hspace*{ 0.9 cm} \(\bullet\) attributes, \\
\hspace*{ 0.9 cm} \(\bullet\) values, \\
\hspace*{ 0.9 cm} \(\bullet\) agents, \\
\hspace*{ 0.9 cm} \(\bullet\)  perspectives, \\
\hspace*{ 0.9 cm} \(\bullet\)  contexts, \\
\hspace*{ 0.9 cm} \(\bullet\)  views.

{\bf Objects} are e.g. components, parts or subsystems  of a manufactured 
artifact, or even the whole technologies, depending on the resolution level 
of a specific snapshot (view) within the FRASMod. 

{\bf Attributes} are characterized  by {\it linguistic 
descriptors} and/or  physical or virtual measurement scales.   
Examples of linguistic descriptors are: 
small\_processing\_windows, high\_temperature, 
good\_lubricity, \\ 
low\_variance\_in\_raw\_material\_costs, etc. 
Examples of attributes that can also be characterized by {\it measurable} 
physical or fiscal parameters are:\\ 
temperature, lubricity, cost\_reducing\_potential, potential\_investment, 
cost, etc.

{\bf Interactions} (special kinds of relations), for example:\\  
REL\_{1,5}: \  \  \  low\_variance\_in\_raw\_material\_costs \(\longrightarrow\) 
common/standard material/alloy system \\
REL\_{2,3}: \  \  \  \  good\_processing\_control \(\longrightarrow\) 
low\_raw\_material\_cost \\ 
REL\_3,7: low\_raw\_material\_cost \(\longrightarrow\) 
common/standard\_material/alloy system.

{\bf Values.} \  \  Values are  assigned to linguistic variables or numerical 
variables, to express the magnitude of a physical or fiscal parameters 
of the attributes, or the truth-value (i.e. the degree to which an  
object possesses an attribute).

{\bf Perspectives.} \  \  An object or a family of objects can be 
evaluated within different perspectives. For example, an LPT cover 
plate can be evaluated from the perspective of an engineer, or from 
the perspective of a business analyst performing value analysis 
of the part, or from the perspective of an accountant. Each perspective 
may employ attributes that are different from the attributes of a 
different perspective for the same object. Some attributes may, however, 
be shared by different perspectives.

{\bf Contexts.} \  \   Each object or family of objects can appear in 
several different contexts. For example an LPT cover plate may 
appear e.g. in context of ingot process, forging process, extrusion process, 
or other processes. 
                       
{\bf Views.} \  \   Even in one particular perspective or context, 
different experts may assess the objects and situations in which 
objects appear differently. These differences of views of different 
experts can be captured by repertory grids and compared by relational 
methods using algorithms provided by TRYSIS. 

{\bf Agents.} \  \  In the context of this project, agents are the 
observers (e.g. engineers or accountants) assessing the degree to which 
an attribute is possessed by an object. For example, in [R4], [R8], [R10] 
describing the evaluation of an LPT cover plate the observers were engineers 
evaluating to what degree various attributes can be assigned to the 
LPT plate.

\subsection{The Role of FRASMod Knowledge Representation Scheme in
Affordability Studies} 

FRASMod has formed the backbone of the whole project, making it possible to
link data collection, data analysis and evaluation  in a unified framework
that is computer representable.
It also allows us  to represent  the cost and  
performance targets and other design criteria in the same framework. 
In this unified framework, we also perform analysis of uncertainty,  
 fuzzy indeterminacy and evaluation of consistency of data and knowledge. 

We cannot discuss here all the uses the  FRASMod scheme was put to in this
project in its entirety. Here  we focus on relational data
analysis and representation.

\subsection{Application of Fuzzy Relational Methods in Evaluation of
Affordability of a Manufacturing Process} \ \ 

This section is concerned with the work related to Objective 1 listed
above in Sec. 1.3. 

We 
have developed methods for Knowledge elicitation and relational 
representation \cite{89.1},\cite{92.7},\cite{91.15} 
of the substantive knowledge  (concepts, linguistic descriptors, 
physically measurable parameters  and 
interactions) that are relevant to the affordability analysis and 
prediction and are applicable not only to {\it technical} but also to 
{\it  human} and {\it organizational} subsystems  of a total production 
system. 

We have designed a set of repertory grids to capture the expertise 
of engineers \cite{96.4},\cite{97.1},\cite{97.4} which is one of the most
important sources of information  in the situations where no historical 
data on the manufactured product are available. 
Repertory grids utilize verbal descriptors, thus  making it possible to assign 
different levels of accuracy, precision, or certainty to each part 
and process, such as cost, material input, or processing condition 
\cite{97.4}.
Pratt \& Whitney engineers found the repertory grids not difficult 
to comprehend and quick to fill in. This is important in a busy 
industrial environment. 

The data has been collected and analysis performed so far at 
{\it three different resolution levels}: 

\noindent                  
{\sf (1) the level of component parts of an aircraft jet engine}: 
analyzing e.g. a $\gamma$-titanium Low Pressure Turbine (LPT) 
cover plate  [R4],[R5],[R8],[R10]. Comparison with other parts 
(titanium rings) and materials (e.g. nickel) is in progress.  \\
{\sf (2) The level of integration components into a subsystem:}    
developing a fuzzy  model for computing
interval bounds of the cost of the subsystem as a function of parts and
values of the process attributes.  \\
{\sf (3) the level of cost estimation of competing technologies:} [R5],[R16]. 
This also provides interval bounds. (See Sec. 4  below.)

In general, affordability modeling involves a variety of contexts and 
resolution levels, e.g. level of parts, processes, assembled artifacts, 
cost/performance tradeoffs, business practices, etc. 
(See Fig. 1 above). 

For relating information concerning the structures of 
these different resolution levels, we have developed the technique of 
generalized morphisms [R3], [R7]. This makes it possible integrate 
separate models of different resolution levels into one 
multi-resolution global model, interrelating the relevant cost related 
features. 
Generalized morphisms (GMorphs)are also important for ensuring the correctness 
of scale measurements by repertory grids. Mathematically, GMorphs
\cite{86.2} are generalizations of
homomorphisms that play an important role in the theory of measurement 
\cite{turksen.fss40}.
\\

Collecting engineering data concerning various manufacturing processes,
parts, subsystems and products  is usually done by physical measurements 
of such physical entities that serve as {\it cost drivers}. 
Because one of our concerns is to deal with affordability
issues when such ``hard" data are not available, we shall, however, present 
here the results of relational analysis on data and knowledge elicited 
by questioning engineers. Here, instead of physical measurement devices 
we use psychometric tools invented by behavioral scientists called {\it
repertory grids} (RPG). 
Our relational analysis can be used to analyze data (e.g. process parameters) 
collected by physical measurements as well as data obtained by knowledge
elicitation from human experts.

A substantial effort in this project was devoted to exploratory knowledge 
elicitation that made it possible to develop such grids for problems 
relevant to the problem area of our industrial partner -- integrating 
affordability into IPPD environment.  Here is a brief summary of how 
RPGrids have been developed and utilized: 

The entities of the processes were identified by exploratory knowledge
elicitation and the cost drivers called process constructs ($c_i$) for each
process selected. Using these results {\it repertory grids} (RPG) with
bi-polar
constructs were developed (Fig. 2 gives an example of a RPG). 
These RPGs were used to elicit information about relationships
of process constructs by presenting these to Pratt\& Whitney engineers.

By converting the grids to relational matrices and processing these by the 
TRYSIS system  tests for various relational properties were performed. 
The computational tests for this purpose are based on  
BK-Products of relations and Fast Fuzzy Relational 
Algorithms \cite{91.15}. These tests make explicit 
relational structures and properties intrinsically contained in data.
Testing the fuzzy relational structure for various relational properties
allows us to discover dependencies, hierarchies, 
similarities, and equivalences of
the attributes characterizing technological processes and manufactured
artifacts  in their relationship to costs and performance. 

The  example of the ingot process
shows dependences of the process constructs/cost-drivers represented as
Hasse diagrams (see Fig. 3, Fig. 4 and Fig. 7) that are standard way of 
representing preorders.

\section{LPT Cover Plate Relational Analysis: A demonstrator 
project} 

For the demonstration of our relational analysis method, our 
industrial partner selected a jet engine component, a {\it Low Pressure 
Turbine (LPT) Cover Plate}\cite{96.4,97.1}. \\

Selecting  a set of objects (e.g. engine parts), using repertory grids 
we have isolated technological attributes of these objects which are 
relevant to the cost and expressed as a fuzzy relational structures [R3]. 
Testing these structure for various relational properties yields  
dependencies, hierarchies, similarities, and equivalences of 
the attributes significant with respect to cost [R4]. Carry out this 
we had to develop the appropriate methodology.

Using a LPT cover plate as the  appropriate object for a demonstration of
our techniques, we have: \\[-0.5 cm]

\begin{quote}
\noindent 
$\bullet$ Performed exploratory knowledge elicitation that resulted in selecting cost
drivers for evaluating the affordability of LPT cover plate in all 5 
processes involved in its manufacturing. \\
\noindent 
$\bullet$
Designed a set of Repertory grids for collecting data on LPT cover plate
from engineers. \\
\noindent 
$\bullet$
Performed a set of experiments eliciting the values of process parameters
for the LPT cover plate. (See  description of the three 
scenarios for  the use of repertory grids and results 
in Sec. 3.2 below.)\\
\noindent 
$\bullet$
Developed a method for comparison of different but similar parts with
respect to process parameters and other attributes 
(see Sec. 3.2.2 below.) \\[-0.5 cm] 
\end{quote}

\subsection{The objective of the relational analysis of the LPT Cover Plate}

A  {\it Low Pressure Turbine (LPT) Cover Plate} is to be manufactured, using
new material, namely, gamma titanium. Prior to any production
characterization, the part is to be costed out, using the expert knowledge 
concerning manufacturing processes and  available cost estimation
that is available for other small gamma titanium parts.    

This is a part with the limited characterization data in  
processes with little manufacturing base, for which 
only very limited empirical data are available. Hence 
elicitation of knowledge of human experts and further fuzzy 
relational extrapolation are necessary.  
 
\begin{figure} [htbp]
\centering 
{\large \caption{ \bf Repertory Grid Analyzer (RPGA)}}
{\bf A sample of input data for RPGA: LPT Cover Plate (Ingot process)}

\vspace{.1in}
\centering
\small 
\hspace{-.5in} 

\begin{tabular}{|c||c|c||c|c|c||c||c|c|c||c|c||}\hline
  \multicolumn{1}{|c||}{}
& \multicolumn{2}{c||}{Primay pole}
& \multicolumn{1}{c|}{}
& \multicolumn{1}{c|}{}
& \multicolumn{1}{c||}{}
& \multicolumn{1}{c||}{}
& \multicolumn{1}{c|}{}
& \multicolumn{1}{c|}{}
& \multicolumn{1}{c||}{}
& \multicolumn{2}{c||}{Secondary pole} \\ \cline{2-3}\cline{11-11}
  \multicolumn{1}{|c||}{Pr.}
& \multicolumn{1}{c|}{description}
& \multicolumn{1}{c||}{range}
& \multicolumn{1}{c|}{3}
& \multicolumn{1}{c|}{2}
& \multicolumn{1}{c||}{1}
& \multicolumn{1}{c||}{0}
& \multicolumn{1}{c|}{-1}
& \multicolumn{1}{c|}{-2}
& \multicolumn{1}{c||}{-3}
& \multicolumn{1}{c|}{description}
& \multicolumn{1}{c||}{range}\\ \hline
  & Very low                    &  	& & & & & & & & Fairly high    		& 	  \\
1 & \% of total cost            & $15\%$ & & & & & & $\surd$ & &  \% of total cost      & $30\%$       \\ \hline
  & Low                         &	& & & & & & & & High                    &	          \\
2 & raw material costs          & \$$10/lb$ & & & & & $\surd$  & & & raw material costs    & \$$40/lb$ 	  \\ \hline
  & Low variability             &	& & & & & & & & High variability        &	       \\
3 & in raw material costs       & $\pm5\%$ & & & & $\surd$  & & & & in raw material costs & $\pm20\%$ 	      \\ \hline
  & Good process control        &	& & & & & & & & Poor process control    & 	          \\
4 & of raw materials            & $C_{pk} \geq 1.3$	& & & & $\surd$  & & & & of raw materials        & $C_{pk} \geq 0$ 	      \\ \hline
  & Standard 		        & $24", 28",$  & & & & & & & & Non-standard 	&	   \\	
5 & size of ingot               & $30", 32"$   & & & & & $\surd$  & & & size of ingot    & 	\\ \hline	
  & Small                       &	& & & & & & & & Large                   & 	      \\
6 & ingot weight                & $600 lb$ & & & & $\surd$  & & & & ingot weight         & $2500 lb$      \\ \hline
  & Short                       & $2$	   & & & & & & & & Long                    & $12$	 	    \\
7 & raw material lead time      & $months$ & & & & & $\surd$  & & & raw material lead time  & $months$ 	\\ \hline
  & Common/standard             &	& & & & & & & & New                     &		\\
8 & material/alloy system       &	& & & & & $\surd$  & & & material/alloy system   & 	     \\ \hline
  & Small variation             &	& & & & & & & & Large variation         & 	\\	
9 & in material properties      &	& & & & $\surd$  & & & & in material properties  &	\\ \hline	
  & Small                       &	& & & & & & & & Large                   &	\\	
10 & numbers of defects         & 	& & & & $\surd$  & & & & numbers of defects      &	\\ \hline	
  &                             &	& & & & & & & &                         & 	     \\
11 & 100 \% yield               & $100\%$ & & & & $\surd$  & & & & $25 \%$ yield         & $25\%$	     \\ \hline
  &  Low                        &	& & & & & & & & High                    & 	  \\
13 & cracking probability       & $5\%$	& & & & & & $\surd$  & & cracking probability    & $50\%$  \\ \hline
\end{tabular} 

\vspace{.1in}
\end{figure}

\vspace*{.2in}

\begin{figure} [htbp]
\centering
{\large \caption{Ingot}}
 {\bf A sample of output results from RPGA package\\}
\vspace*{.1in}
 \includegraphics{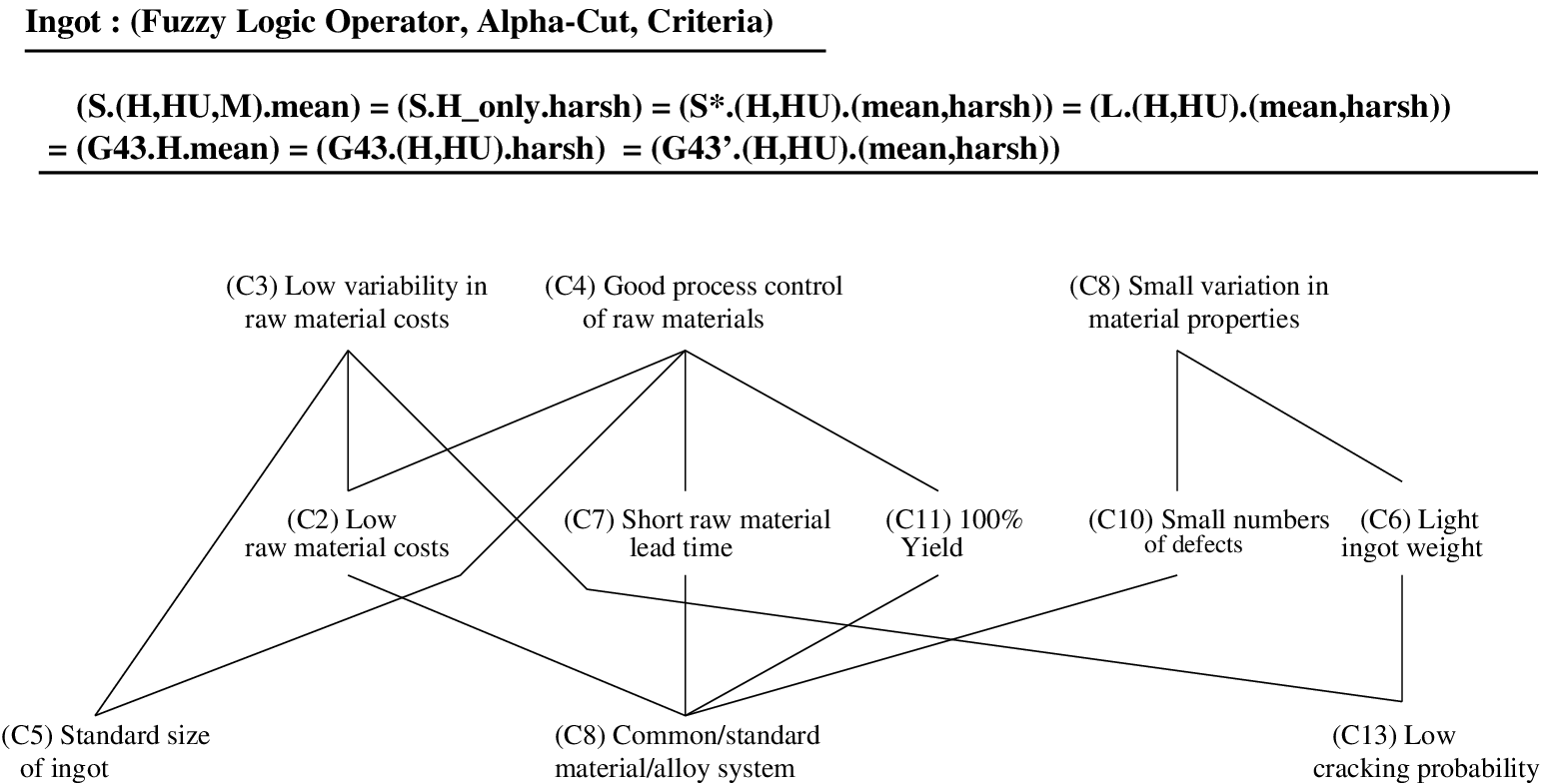}
\end{figure}

The meaning of relational sorts (semiotic descriptors) in the  Fuzzy 
Relational Affordability Systemic Model (FRASMod) is as follows. 

\noindent
{\bf {\sf Object:}} \  \  Low Pressure Turbine (LPT) Cover Plate.\\ 
{\bf {\sf Perspective:}} \  \ Dependency of the cost on the product -- 
process  relationship. \\
{\bf {\sf Contexts:}} \  \ Five processes during the manufacturing 
of the LPT Cover Plate, namely: ingot process, forging, extrusion, 
heat treatment, machining. \\ 
{\bf \sf Agents.} \  \  Agents are the respondents -- 5 Pratt \& 
Whitney engineers filling in the repertory grids, thus providing 
information about process' attributes of the nickel or 
$\gamma$-titanium LPT Cover Plate. \\
{\bf {\sf Attributes:}} \  \   Process' entities selected 
as cost drivers, represented as bipolar constructs, in the repertory 
grids that were presented to respondents -- engineers. \\
{\bf {\sf The Evaluative Task Structure:}} \  \  
A number of elicitation and evaluative schemes {\it (Scenarios)} can 
be formulated, capturing inter-process dependences, inter-observer 
dependences, etc.\\[-0.4 cm]

\subsection{Knowledge Elicitation and Data Analysis}

There is a number of problems that can be solved by relational analysis 
of information obtained by repertory grids. In this section we outline 3
different scenario for evaluation of an LPT-plate, namely: 

\begin{enumerate}
\item Discovering dependency structures of cost drivers.
\item Identification of characteristic similarities and differences between
parts made of different materials. 
\item Discovering interprocess differences between the meaning of cost drivers. 
\end{enumerate}

\subsubsection{SCENARIO 1: Discovering dependency structures of cost
drivers}

1 object  (LPT cover plate) and a group of respondents (5 engineers).  
Each respondent-engineer has assessed the object independently 
in the five processes involved in manufacturing of the part. 
The aim here is to find the dependences between process cost drivers, 
as well as the inter respondent consistency. 

The resulting Hasse diagram computed from the RPGs (Fig.2) of the {\it ingot 
process} is shown in Fig. 3.                       
Also the dependences  between the judgments of engineers 
have been obtained.

From the Hasse diagrams computed for all the processes 
the necessary and possible fuzzy dependences have been
derived. 

We have to distinguish {it necessary} from {\it possible} 
dependences.  This follows from the logic theory of BK-products 
and Fast Fuzzy Relational algorithms by which the Hasse diagrams 
are computed. 

Let us briefly look at a sample of such dependences as 
they appear in the process of {\it machining}. Their Hasse diagram 
appears in Fig. 4. The verbal statement {\it {\sf x} is necessarily   
dependent on {\sf y and z}} we abbreviate by 
$x \stackrel{ \Box}{\Longrightarrow} y$ \& $z$. 
Similarly, $ {x \not \Longrightarrow} y$ reads {\it x is 
independent of y}.  $x \stackrel{\Diamond}{\Longrightarrow} y \vee z$  reads
{\it {\sf x} is possibly dependent on {\sf y or z}}.  For example, from
Fig. 4 we can read the following. \\

$\bullet$ Necessary Dependencies of the Process 
Parameter $C_8$ \\[-1.1 cm]
                                                                                    
\begin{tabbing}
aaa \= blah \= blah2blah3blah4blah5 \= blah5bb \= blah \= blah6 \= \kill \\ 
\> $C_8$ : \> Machining  \> $\stackrel{\Large \Box}{\Longrightarrow}$    \> $C_2$ : \> Part size \&  \\
\>      \> distortion/warpage    \>  \> $C_4$ :   \> Material machinability \&  \\                      
\>      \>      \>     \> $C_{11}$ : \> Machining data \& \\
\>      \>             \>    \>
$C_{14}$ : \> Post-machining inspection \\[-0.7 cm]
\end{tabbing}

 $\bullet$ Non-Dependencies of the Process Parameter $C_8$:\\
$C_1$: Init. part variability ${\not \Longrightarrow}$
 $C_8$: Machining distortion $\vee$ $C_2$:  Part size $\vee$ 
$C_{14}$: Post-machining inspection \\

Some of these dependencies and independencies appear fairly 
obvious to an engineer with some experience other dependences 
are less obvious but can be validated.   
The essential point here to realize is that all this inference 
 has been obtained computationally from the repertory grids, each 
of which was filled  by a different expert within a few minutes.

\begin{figure} [htbp]

\hspace{1in} 
\includegraphics{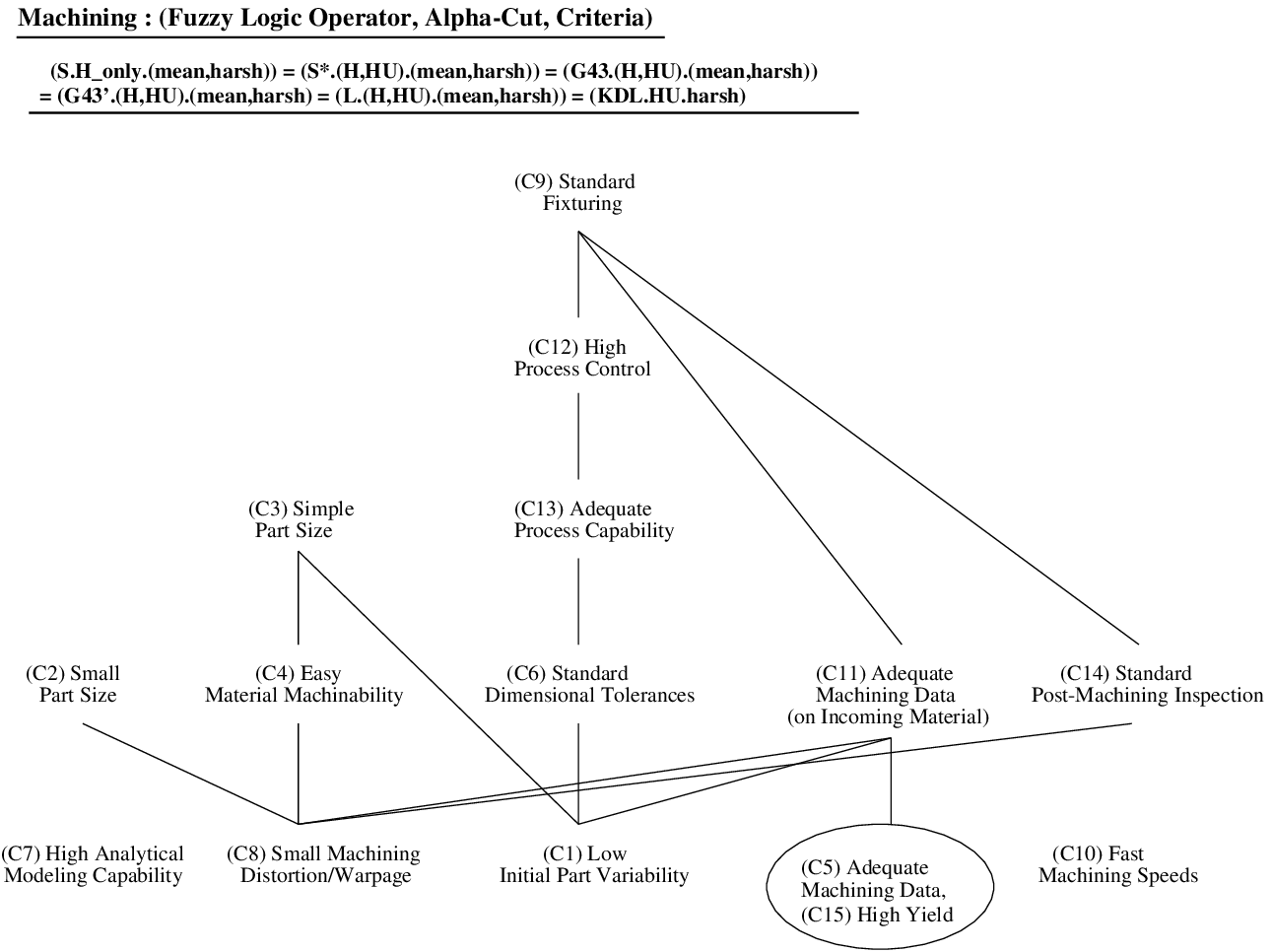}
\caption{Hasse Diagrams of Machining}
\end{figure}

\subsubsection{SCENARIO 2: Similarities and differences of 
parts made of different materials} 

Determining characteristic similarities and differences between
parts made of different materials may involve one or more respondents (engineers)  and a collection of objects 
(e.g. different LPT parts). 
Here, the aim is to detect {\sf characteristic similarities} and {\sf differences} 
between distinct objects. 

We have chosen to compare the LPT cover plates made of two
different materials, namely {\it nickel} and {\it $\gamma$-titanium} 
in all 5 manufacturing processes. The results are summarized in Table 1. 
Degrees of similarity were computed by the fuzzy logic using the  
the fuzzy equivalence operator based on the {\L}ukasiewicz implication 
operator. Degrees of difference were computed by the operator dual to 
the fuzzy equivalence operator. 

The  differences on a relative scale are plotted as bar-charts 
(for a sample see Fig. 5). The classivalence classes relating nickel 
and $\gamma$-titanium data were also computed [R19].

\begin{figure} [htb]
 \includegraphics{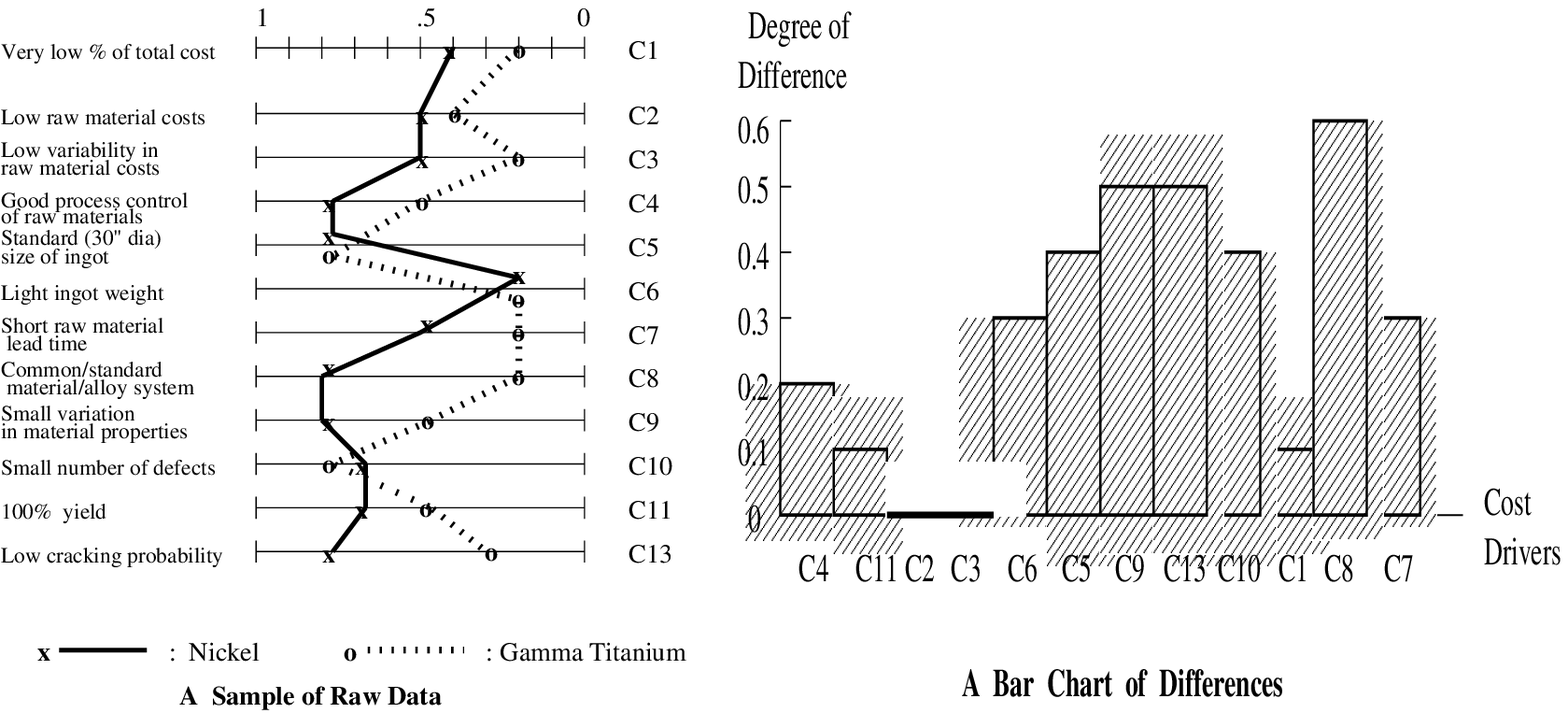}
\vspace{.2in}
\caption{}
\end{figure}

We have seen in Scenario 1 that testing for preorders reveals possible
dependences of process entities (in particular those selected as the cost
drivers). Knowing mutual dependencies allows for identifying these
interrelationships. In the Scenario 2 another kind of generalized 
equivalence, namely {\it classivalence} \cite{86.2} appears very useful. 
One may ask what classivalence, a generalized ``equivalence" of two 
different sets is. 

In general, equivalence may appear in a relation from a set to itself. 
Classivalence, 
related to bifunctionality can appear when two {\it different sets} are 
related by a relation.

Equivalence and classivalence classes identified in the data of Scenario 2 
provide the information as to which process entities may 
have equivalent effect, hence can be treated as interchangeable in their 
impact on the other portions of affordability models.  
More detailed explanation of classivalence appears in [R19].

\begin{table} [htbp]

\hspace{-.5in} \caption{Summarization of differences of cost drivers of a LPT cover plate: Nickel material vs. $\gamma$=titanium}
\vspace{.1in}

{\footnotesize \begin{tabular}{|c||c|c|c|c|c||c|}\hline
  \multicolumn{1}{|c||}{}
& \multicolumn{5}{c|}{Processes} 
& \multicolumn{1}{|c|}{} \\ \cline{2-6}
  \multicolumn{1}{|c||}{Measures}
& \multicolumn{1}{c|}{Ingot}
& \multicolumn{1}{c|}{Extrusion}
& \multicolumn{1}{c|}{Forging}
& \multicolumn{1}{c|}{Heat Treatment}
& \multicolumn{1}{c||}{Machining}
& \multicolumn{1}{c|}{LPT global} \\ \hline \hline
			&	&	&	&	&	&	\\
\# of Cost Drivers	& 12	& 19	& 27	& 13	& 15	& 86	\\ \hline
			&	&	&	&	&	&	\\
Mean Difference		& 0.28	& 0.185	& 0.24	& 0.23	& 0.2	& 0.23	\\ \hline
Max Value		&	&	&	&	&	&	\\
of Difference & $<0, 0.5>$ & $<0, 0.4>$ & $<0, 0.6>$ & $<0, 0.4>$ & $<0, 0.5>$ & $<0, 0.48>$ \\ \hline
\# of Cost Drivers 	&	&	&	&	&	&	\\
in $<max, max-10\%>$	& 3	& 9	& 4	& 7	& 5	& 28	\\ \hline 
\# of Cost Drivers 	&	&	&	&	&	&	\\
in $<min, min+10\%>$	& 4	& 8	& 7	& 4	& 5	& 28	\\ \hline
\# of Cost Drivers 	&	&	&	&	&	&	\\
with Similarity $\geq 70\%$ 	& 7	& 15	& 17	& 9	& 8	& 56	\\ \hline
Percentage of Cost Drivers	&	&	&	&	& 	& \\
with Similarity $\geq 70\%$	& 58\%	& 80\%	& 63\%	& 69\%	& 53\%	& 65\%	\\  \hline
\end{tabular}
}
\end{table} 

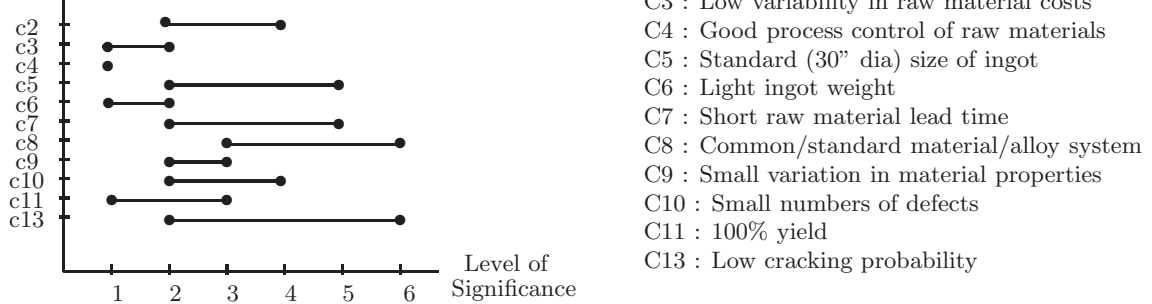
\begin{figure}  [htbp] 
\setlength{\unitlength}{0.00066700in}%
\begingroup\makeatletter\ifx\SetFigFont\undefined
\def\x#1#2#3#4#5#6#7\relax{\def\x{#1#2#3#4#5#6}}%
\expandafter\x\fmtname xxxxxx\relax \def\y{splain}%
\ifx\x\y   
\gdef\SetFigFont#1#2#3{%
  \ifnum #1<17\tiny\else \ifnum #1<20\small\else
  \ifnum #1<24\normalsize\else \ifnum #1<29\large\else
  \ifnum #1<34\Large\else \ifnum #1<41\LARGE\else
     \huge\fi\fi\fi\fi\fi\fi
  \csname #3\endcsname}%
\else
\gdef\SetFigFont#1#2#3{\begingroup
  \count@#1\relax \ifnum 25<\count@\count@25\fi
  \def\x{\endgroup\@setsize\SetFigFont{#2pt}}%
  \expandafter\x
    \csname \romannumeral\the\count@ pt\expandafter\endcsname
    \csname @\romannumeral\the\count@ pt\endcsname
  \csname #3\endcsname}%
\fi
\fi\endgroup
\begin{picture}(4050,2065)(-451,-8693)
\thicklines
\put(1471,-6586){\circle*{90}}
\put(1021,-6781){\circle*{90}}
\put(1021,-6931){\circle*{90}}
\put(1501,-7381){\circle*{90}}
\put(1951,-7531){\circle*{90}}
\put(3301,-7531){\circle*{90}}
\put(1951,-7681){\circle*{90}}
\put(1501,-7681){\circle*{90}}
\put(1501,-7831){\circle*{90}}
\put(2371,-7831){\circle*{90}}
\put(1051,-7981){\circle*{90}}
\put(1951,-7981){\circle*{90}}
\put(3301,-8131){\circle*{90}}
\put(2371,-6611){\circle*{90}}
\put(1501,-7223){\circle*{90}}
\put(1501,-6781){\circle*{90}}
\put(1026,-7223){\circle*{90}}
\put(2821,-7381){\circle*{90}}
\put(1501,-7081){\circle*{90}}
\put(2821,-7081){\circle*{90}}
\put(1501,-8131){\circle*{90}}
\put(676,-6136){\line( 0,-1){2400}}
\put(2851,-8486){\line( 0,-1){ 75}}
\put(646,-6761){\line( 1, 0){ 75}}
\put(646,-6911){\line( 1, 0){ 75}}
\put(646,-7061){\line( 1, 0){ 75}}
\put(646,-7211){\line( 1, 0){ 75}}
\put(646,-7361){\line( 1, 0){ 75}}
\put(646,-7511){\line( 1, 0){ 75}}
\put(646,-7661){\line( 1, 0){ 75}}
\put(646,-7811){\line( 1, 0){ 75}}
\put(646,-8111){\line( 1, 0){ 75}}
\put(646,-7961){\line( 1, 0){ 75}}
\put(1501,-6611){\line( 1, 0){900}}
\put(976,-6781){\line( 1, 0){525}}
\put(1501,-7081){\line( 1, 0){1350}}
\put(1501,-7381){\line( 1, 0){1320}}
\put(1501,-7681){\line( 1, 0){450}}
\put(1501,-7831){\line( 1, 0){900}}
\put(1051,-7981){\line( 1, 0){900}}
\put(1501,-8141){\line( 1, 0){1800}}
\put(1951,-7541){\line( 1, 0){1350}}
\put(2401,-8486){\line( 0,-1){ 75}}
\put(646,-6611){\line( 1, 0){ 75}}
\put(1051,-8486){\line( 0,-1){ 75}}
\put(1951,-8486){\line( 0,-1){ 75}}
\put(676,-8536){\line( 1, 0){2925}}
\put(3301,-8486){\line( 0,-1){ 75}}
\put(1501,-8486){\line( 0,-1){ 75}}
\put(1026,-7223){\line( 1, 0){475}}
\put(1801,-5936){\makebox(0,0)[lb]{\smash{\SetFigFont{10}{12.0}{bf}$\alpha = .17$}}}
\put(3701,-8731){\makebox(0,0)[lb]{\smash{\SetFigFont{9}{10.8}{rm}Significance}}}
\put(3801,-8531){\makebox(0,0)[lb]{\smash{\SetFigFont{9}{10.8}{rm}Level of}}}
\put(1051,-8766){\makebox(0,0)[lb]{\smash{\SetFigFont{9}{10.8}{rm}1}}}
\put(1501,-8766){\makebox(0,0)[lb]{\smash{\SetFigFont{9}{10.8}{rm}2}}}
\put(1951,-8766){\makebox(0,0)[lb]{\smash{\SetFigFont{9}{10.8}{rm}3}}}
\put(2401,-8766){\makebox(0,0)[lb]{\smash{\SetFigFont{9}{10.8}{rm}4}}}
\put(2851,-8766){\makebox(0,0)[lb]{\smash{\SetFigFont{9}{10.8}{rm}5}}}
\put(3326,-8766){\makebox(0,0)[lb]{\smash{\SetFigFont{9}{10.8}{rm}6}}}
\put(301,-6836){\makebox(0,0)[lb]{\smash{\SetFigFont{9}{10.8}{rm}c3}}}
\put(301,-6986){\makebox(0,0)[lb]{\smash{\SetFigFont{9}{10.8}{rm}c4}}}
\put(301,-7136){\makebox(0,0)[lb]{\smash{\SetFigFont{9}{10.8}{rm}c5}}}
\put(301,-7286){\makebox(0,0)[lb]{\smash{\SetFigFont{9}{10.8}{rm}c6}}}
\put(301,-7436){\makebox(0,0)[lb]{\smash{\SetFigFont{9}{10.8}{rm}c7}}}
\put(301,-7586){\makebox(0,0)[lb]{\smash{\SetFigFont{9}{10.8}{rm}c8}}}
\put(301,-7736){\makebox(0,0)[lb]{\smash{\SetFigFont{9}{10.8}{rm}c9}}}
\put(251,-8036){\makebox(0,0)[lb]{\smash{\SetFigFont{9}{10.8}{rm}c11}}}
\put(251,-8186){\makebox(0,0)[lb]{\smash{\SetFigFont{9}{10.8}{rm}c13}}}
\put(251,-7886){\makebox(0,0)[lb]{\smash{\SetFigFont{9}{10.8}{rm}c10}}}
\put(151,-6066){\makebox(0,0)[lb]{\smash{\SetFigFont{9}{10.8}{rm}Cost Driver}}}
\put(301,-6686){\makebox(0,0)[lb]{\smash{\SetFigFont{9}{10.8}{rm}c2}}}
\put(5201,-6261){\makebox(0,0)[lb]{\smash{\SetFigFont{9}{10.8}{rm}C2  : Low raw material cost}}}
\put(5201,-6486){\makebox(0,0)[lb]{\smash{\SetFigFont{9}{10.8}{rm}C3  : Low variability in raw material costs}}}
\put(5201,-6711){\makebox(0,0)[lb]{\smash{\SetFigFont{9}{10.8}{rm}C4  : Good process control of raw materials}}}
\put(5201,-6936){\makebox(0,0)[lb]{\smash{\SetFigFont{9}{10.8}{rm}C5  : Standard (30" dia) size of ingot}}}
\put(5201,-7161){\makebox(0,0)[lb]{\smash{\SetFigFont{9}{10.8}{rm}C6  : Light ingot weight}}}
\put(5201,-7386){\makebox(0,0)[lb]{\smash{\SetFigFont{9}{10.8}{rm}C7  : Short raw material lead time}}}
\put(5201,-7611){\makebox(0,0)[lb]{\smash{\SetFigFont{9}{10.8}{rm}C8  : Common/standard material/alloy system}}}
\put(5201,-7836){\makebox(0,0)[lb]{\smash{\SetFigFont{9}{10.8}{rm}C9  : Small variation in material properties}}}
\put(5201,-8061){\makebox(0,0)[lb]{\smash{\SetFigFont{9}{10.8}{rm}C10 : Small numbers of defects}}}
\put(5201,-8286){\makebox(0,0)[lb]{\smash{\SetFigFont{9}{10.8}{rm}C11 : 100\% yield}}}
\put(5201,-8511){\makebox(0,0)[lb]{\smash{\SetFigFont{9}{10.8}{rm}C13 : Low cracking probability}}}

\end{picture}
\vspace{.1in}
\caption{Interval ranking of cost drivers for LPT cover plate made of $\gamma$-titanium in process Ingot}
\end{figure}

\subsubsection{SCENARIO 3: Interprocess differences between the meaning of
cost drivers}

This involves one or more respondents (engineers), 1 object, several 
situations or processes in which the object may appear. 
In this scenario, the primary 
goal is to detect similarities and dependences between process attributes 
of different processes. See Fig. 7. for results of comparing 
cost drivers of the extrusion process with the cost drivers in  the forging 
for the $\gamma$-titanium LPT cover plate.  

Forging and extrusion are interesting from the point of view of 
investigating the effect of context on the meaning of process 
attributes. As we can see from Table 1 (which also shows the number of cost 
drivers in each process) {\it forging} has {\it 27 cost drivers} 
(RPG bipolar constructs), while {\it extrusion} has only 
{\it 19 cost drivers}. Out of this number, 9 cost drivers 
(RPG bipolar constructs) are overlapping:  
they have the same linguistic labels -- names (see Fig. 7 for 
their names). In each process, although having the same name 
denoting the same general concept, they may 
have different minimum and maximum ranges assigned. The question 
then the arises: 
do the overlapping bipolar constructs interact in each context 
in the same way? This is an empirical question answer to which 
cane be provided amongst other things) by the 
experimental arrangement 
of Scenario 3.  

We can see that the Hasse diagrams capturing the ordering 
(it is a pre-order) of cost drivers having the same linguistic label 
``name" are different. Hence their meaning is different, because the 
contexts, namely processes are different.

Comparing the Hasse diagrams for  forging and extrusion 
in Fig. 7  we can see that only the constructs  $c_2$: {\it  process 
window} $c_7$: {\it Tooling} have a dependency in common 
$c_2 \Longrightarrow c_7$
This link appears in both extrusion and forging.

This however does not mean that a specific list of repertory grid constructs
having identical name have the same meaning in two different contexts. 

Looking at equivalences in two different contexts we can see the 
following. Fig. 7 shows that in the context of {\it extrusion}, 
semiotic descriptors 
$c_2$: {\it large process window} and $c_9$: {\it long die life}  
lie within the equivalence class, while in the context of {\it forging} 
$c_2$: {\it Large process 
window} is equivalent with $c_5$: {\it air furnace atmosphere}. 
The equivalence of $c_2$: {\it large process window } and 
$c_9: ${\it long die life}, however, {\bf does not hold} in the context 
of {\it forging} despite of the fact that it holds for extrusion.

Data can also be analyzed taking the negative side of bi-polar
PRG constructs. The preorder depicted in Figure 7 on the left shows the 
property of {\it contrapositive symmetry}\footnote{A logic proposition is contrapositive
if $a \rightarrow b = \neg b \rightarrow \neg a$.}. 

We have e.g. 
$c_3 \Rightarrow  c_4 \Rightarrow \{c_2, c_5\}  \Rightarrow c_7$
$=  \neg c_7 \Rightarrow \neg \{c_2, c_5\}  \Rightarrow  \neg c_4 \Rightarrow  \neg c_3$.

The Hasse diagrams on the right side, however,  do not have contrapositive 
property. 
Hence, the presence or absence of contrapositivity is an is important 
characteristic of data that ought to be always tested.  

\subsection{Integration of Perspectives and Resolution Levels  of Relational Models and Summarization of Data}

In general, affordability modeling involves a variety of contexts and 
resolution levels, e.g. level of parts, processes, assembled artifacts, 
cost/performance tradeoffs, business practices, etc. 
(See Fig. 1 above). 
In terms of fuzzy relational models we say that 
each resolution level represents different granularity 
\cite{96.10},\cite{97.11},\cite{zad.col2} pp. 433-448 
and  \cite{zad.fss97gran}. These different perspectives and models at different resolution levels have to be apprpriately integrated.

\newpage
 
\begin{table} [htbp]
\small
\hspace{.8in} \begin{tabular}{|c|l||c|l|}  \hline
  \multicolumn{2}{|c||}{Positive Semiotic Descriptors}
& \multicolumn{2}{c|}{Negative Semiotic Descriptors} \\ \hline
  \multicolumn{1}{|c|}{Symbol}
& \multicolumn{1}{c||}{Meaning}
& \multicolumn{1}{c|}{Symbol}
& \multicolumn{1}{c|}{Meaning} \\ \hline
C1  
& Capable Analytical Modeling 
& $\overline{C1}$  
&Limited Analytical Modeling \\ 
C2 
& Large Process Window
& $\overline{C2}$   
& Small Process Window \\ 
C3 
& Low Temperature 
& $\overline{C3}$ 
& High Temperature \\ 
C4 
& Good Lubricity 
& $\overline{C4}$ 
& Low(or Difficult) Lubricity \\ 
C5  
&
Air Furnace Atmosphere 
& $\overline{C5}$ 
& Vacuum Furnace Atmosphere \\ 
C6 
& Good Process Control
& $\overline{C6}$ 
& Limited Process Control \\ 
C7 
& Available Tooling
& $\overline{C7}$ 
& New Tooling \\ 
C8 
& Flat Die Shape
& $\overline{C8}$ 
& Shaped Die Shape \\ 
C9 
& Long Die Life
& $\overline{C9}$ 
& Short Die Life \\ \hline
\end{tabular}

\end{table}

\vspace{1in}
\noindent
\begin{figure} [htbp]
\hspace{.1in} \setlength{\unitlength}{0.00058300in}%
\begingroup\makeatletter\ifx\SetFigFont\undefined
\def\x#1#2#3#4#5#6#7\relax{\def\x{#1#2#3#4#5#6}}%
\expandafter\x\fmtname xxxxxx\relax \def\y{splain}%
\ifx\x\y   
\gdef\SetFigFont#1#2#3{%
  \ifnum #1<17\tiny\else \ifnum #1<20\small\else
  \ifnum #1<24\normalsize\else \ifnum #1<29\large\else
  \ifnum #1<34\Large\else \ifnum #1<41\LARGE\else
     \huge\fi\fi\fi\fi\fi\fi
  \csname #3\endcsname}%
\else
\gdef\SetFigFont#1#2#3{\begingroup
  \count@#1\relax \ifnum 25<\count@\count@25\fi
  \def\x{\endgroup\@setsize\SetFigFont{#2pt}}%
  \expandafter\x
    \csname \romannumeral\the\count@ pt\expandafter\endcsname
    \csname @\romannumeral\the\count@ pt\endcsname
  \csname #3\endcsname}%
\fi
\fi\endgroup
\begin{picture}(5109,6510)(304,-7063) 
\thicklines
\put(2076,-2086){\oval(850,450)}
\put(2026,-5436){\oval(890,450)}
\put(226,464){\line( 1, 0){2025}}
\put(126,-2836){\line( 1, 0){2025}}
\put(151,-961){\line( 1, 0){4575}}
\put(201,-4461){\line( 1, 0){4900}}
\put(2026,-4911){\line( 0,-1){300}}
\put(2026,-5686){\line( 0,-1){300}}
\put(2001,-5986){\line(-3, 1){945}}
\put(1995,-6001){\line( 3, 1){1077}}
\put(2026,-6361){\line( 0,-1){300}}
\put(2026,-6661){\line( 4, 1){1058}}
\put(2101,-1486){\line( 0,-1){375}}
\put(2101,-1861){\line( 3, 1){1125}}
\put(2101,-1486){\line( 5,-2){1125}}
\put(2101,-1486){\line(-3,-1){1282}}
\put(826,-1936){\line( 0, 1){450}}
\put(2026,-5211){\line( 4, 1){1058}}
\put(2026,-4936){\line( 1,-1){1087.500}}
\put(751,-1411){\makebox(0,0)[lb]{\smash{\SetFigFont{8}{9.6}{rm}C4}}}
\put(2001,-1411){\makebox(0,0)[lb]{\smash{\SetFigFont{8}{9.6}{rm}C6}}}
\put(751,-2161){\makebox(0,0)[lb]{\smash{\SetFigFont{8}{9.6}{rm}C8}}}
\put(151,-2161){\makebox(0,0)[lb]{\smash{\SetFigFont{8}{9.6}{rm}C1}}}
\put(151,-136){\makebox(0,0)[lb]{\smash{\SetFigFont{8}{9.6}{rm}= (S*,(H,HU),harsh) = (S*,(H,HU),mean)}}}
\put(151,-361){\makebox(0,0)[lb]{\smash{\SetFigFont{8}{9.6}{rm}= (G43,(H,HU),harsh) = (G43, H,mean)}}}
\put(151,-811){\makebox(0,0)[lb]{\smash{\SetFigFont{8}{9.6}{rm}= (\L,(H,HU),harsh) = (\L,(H,HU),mean)}}}
\put(151,-586){\makebox(0,0)[lb]{\smash{\SetFigFont{8}{9.6}{rm}= ($G43^{\prime}$,(H,HU),harsh) = ($G43^{\prime}$,(H,HU),mean)}}}
\put(201,-3211){\makebox(0,0)[lb]{\smash{\SetFigFont{8}{9.6}{rm}   (S,H,harsh) = (S,H,mean)}}}
\put(201,-3436){\makebox(0,0)[lb]{\smash{\SetFigFont{8}{9.6}{rm}= (S*,(H,HU,M),harsh) = (S*,(H,HU,M),mean)}}}
\put(201,-3661){\makebox(0,0)[lb]{\smash{\SetFigFont{8}{9.6}{rm}= (G43,(H,HU),harsh) = (G43,(H,HU),mean)}}}
\put(201,-3886){\makebox(0,0)[lb]{\smash{\SetFigFont{8}{9.6}{rm}= ($G43^{\prime}$,(H,HU),harsh) = ($G43^{\prime}$,(H,HU),mean)}}}
\put(201,-4111){\makebox(0,0)[lb]{\smash{\SetFigFont{8}{9.6}{rm}= (\L,(H,HU),harsh) = (\L,(H,HU),mean)}}}
\put(201,-4336){\makebox(0,0)[lb]{\smash{\SetFigFont{8}{9.6}{rm}= (KDL,HU,harsh)}}}
\put(301,539){\makebox(0,0)[lb]{\smash{\SetFigFont{8}{9.6}{bf}FRS of Extrusion}}}
\put(151, 89){\makebox(0,0)[lb]{\smash{\SetFigFont{8}{9.6}{rm}(S,H,harsh)= (S,(H,HU,M), mean)}}}
\put(3701,-2161){\makebox(0,0)[lb]{\smash{\SetFigFont{8}{9.6}{rm}C3}}}
\put(301,-2761){\makebox(0,0)[lb]{\smash{\SetFigFont{8}{9.6}{bf}FRS of Forging}}}
\put(1976,-4836){\makebox(0,0)[lb]{\smash{\SetFigFont{8}{9.6}{rm}C7}}}
\put(1691,-5511){\makebox(0,0)[lb]{\smash{\SetFigFont{8}{9.6}{rm}C2, C5}}}
\put(1901,-6261){\makebox(0,0)[lb]{\smash{\SetFigFont{8}{9.6}{rm}C4}}}
\put(3026,-5586){\makebox(0,0)[lb]{\smash{\SetFigFont{8}{9.6}{rm}C6}}}
\put(926,-5586){\makebox(0,0)[lb]{\smash{\SetFigFont{8}{9.6}{rm}C1}}}
\put(3026,-6261){\makebox(0,0)[lb]{\smash{\SetFigFont{8}{9.6}{rm}C8}}}
\put(1901,-6836){\makebox(0,0)[lb]{\smash{\SetFigFont{8}{9.6}{rm}C3}}}
\put(3101,-2161){\makebox(0,0)[lb]{\smash{\SetFigFont{8}{9.6}{rm}C5}}}
\put(3101,-1411){\makebox(0,0)[lb]{\smash{\SetFigFont{8}{9.6}{rm}C7}}}
\put(1726,-2161){\makebox(0,0)[lb]{\smash{\SetFigFont{8}{9.6}{rm}C2, C9}}}
\put(3026,-4836){\makebox(0,0)[lb]{\smash{\SetFigFont{8}{9.6}{rm}C9}}}


\end{picture}
\hspace{.1in} \setlength{\unitlength}{0.00058300in}%
\begingroup\makeatletter\ifx\SetFigFont\undefined
\def\x#1#2#3#4#5#6#7\relax{\def\x{#1#2#3#4#5#6}}%
\expandafter\x\fmtname xxxxxx\relax \def\y{splain}%
\ifx\x\y   
\gdef\SetFigFont#1#2#3{%
  \ifnum #1<17\tiny\else \ifnum #1<20\small\else
  \ifnum #1<24\normalsize\else \ifnum #1<29\large\else
  \ifnum #1<34\Large\else \ifnum #1<41\LARGE\else
     \huge\fi\fi\fi\fi\fi\fi
  \csname #3\endcsname}%
\else
\gdef\SetFigFont#1#2#3{\begingroup
  \count@#1\relax \ifnum 25<\count@\count@25\fi
  \def\x{\endgroup\@setsize\SetFigFont{#2pt}}%
  \expandafter\x
    \csname \romannumeral\the\count@ pt\expandafter\endcsname
    \csname @\romannumeral\the\count@ pt\endcsname
  \csname #3\endcsname}%
\fi
\fi\endgroup
\begin{picture}(2609,6510)(154,-7363) 
\thicklines
\put(983,-1588){\oval(1364,504)}
\put(3683,-988){\oval(1364,504)}
\put(983,-5938){\oval(1364,504)}
\put(3669,-5959){\oval(1244,504)}
\put(2604,-6784){\oval(1244,504)}
\put( 76,464){\line( 1, 0){1150}}
\put( 76, 89){\line( 1, 0){1450}}
\put(976,-436){\line( 0,-1){300}}
\put(976,-1036){\line( 0,-1){300}}
\put(976,-436){\line( 3,-2){450}}
\put(976,-1336){\line( 3, 2){450}}
\put(976,-1861){\line( 0,-1){300}}
\put(976,-1861){\line( 5,-3){518}}
\put(3676,-436){\line( 0,-1){300}}
\put(3676,-1261){\line( 0,-1){300}}
\put(3676,-1936){\line( 0,-1){300}}
\put(3676,-2611){\line( 0,-1){300}}
\put(3676,-436){\line(-2,-1){1200}}
\put(2476,-1036){\line( 0,-1){1925}}
\put(3676,-1261){\line( 2,-3){692}}
\put(3676,-2911){\line( 5, 2){685}}
\put( 76,-3736){\line( 1, 0){1050}}
\put( 76,-4186){\line( 1, 0){2150}}
\put(976,-4786){\line( 0,-1){300}}
\put(976,-5386){\line( 0,-1){300}}
\put(976,-6211){\line( 0,-1){300}}
\put(976,-6811){\line( 0,-1){300}}
\put(976,-5686){\line(-2, 1){600}}
\put(976,-6211){\line(-2,-1){600}}
\put(976,-7111){\line(-5, 2){685}}
\put(3676,-4711){\line( 0,-1){300}}
\put(3676,-5386){\line( 0,-1){300}}
\put(3676,-6211){\line( 0,-1){300}}
\put(3676,-4711){\line( 5,-6){952}}
\put(3676,-6211){\line(-4,-1){1058}}
\put(4651,-6211){\line( 0,-1){375}}
\put(4651,-6586){\line(-5, 2){969}}
\put(151,539){\makebox(0,0)[lb]{\smash{\SetFigFont{8}{9.6}{bf}Extrusion}}}
\put(151,164){\makebox(0,0)[lb]{\smash{\SetFigFont{8}{9.6}{rm}(G43, M, mean)}}}
\put(901,-361){\makebox(0,0)[lb]{\smash{\SetFigFont{8}{9.6}{rm}C6}}}
\put(901,-961){\makebox(0,0)[lb]{\smash{\SetFigFont{8}{9.6}{rm}C4}}}
\put(526,-1711){\makebox(0,0)[lb]{\smash{\SetFigFont{8}{9.6}{rm}C2,C7,C9}}}
\put(901,-2386){\makebox(0,0)[lb]{\smash{\SetFigFont{8}{9.6}{rm}C5}}}
\put(1426,-961){\makebox(0,0)[lb]{\smash{\SetFigFont{8}{9.6}{rm}C1}}}
\put(1426,-2386){\makebox(0,0)[lb]{\smash{\SetFigFont{8}{9.6}{rm}C8}}}
\put(301,-2386){\makebox(0,0)[lb]{\smash{\SetFigFont{8}{9.6}{rm}C3}}}
\put(151,-286){\makebox(0,0)[lb]{\smash{\SetFigFont{8}{9.6}{rm}(a)}}}
\put(2401,-361){\makebox(0,0)[lb]{\smash{\SetFigFont{8}{9.6}{rm}(b)}}}
\put(3601,-361){\makebox(0,0)[lb]{\smash{\SetFigFont{8}{9.6}{rm}$\overline{C8}$}}}
\put(3226,-1111){\makebox(0,0)[lb]{\smash{\SetFigFont{8}{9.6}{rm}$\overline{C2}$,$\overline{C7}$,$\overline{C9}$}}}
\put(3601,-1861){\makebox(0,0)[lb]{\smash{\SetFigFont{8}{9.6}{rm}$\overline{C5}$}}}
\put(3601,-2536){\makebox(0,0)[lb]{\smash{\SetFigFont{8}{9.6}{rm}$\overline{C1}$}}}
\put(3601,-3211){\makebox(0,0)[lb]{\smash{\SetFigFont{8}{9.6}{rm}$\overline{C6}$}}}
\put(4201,-2536){\makebox(0,0)[lb]{\smash{\SetFigFont{8}{9.6}{rm}$\overline{C4}$}}}
\put(2401,-3211){\makebox(0,0)[lb]{\smash{\SetFigFont{8}{9.6}{rm}$\overline{C3}$}}}
\put(151,-3661){\makebox(0,0)[lb]{\smash{\SetFigFont{8}{9.6}{bf}Forging}}}
\put(151,-4111){\makebox(0,0)[lb]{\smash{\SetFigFont{8}{9.6}{rm}(G43, M, (harsh,mean))}}}
\put(901,-4711){\makebox(0,0)[lb]{\smash{\SetFigFont{8}{9.6}{rm}C1}}}
\put(901,-5311){\makebox(0,0)[lb]{\smash{\SetFigFont{8}{9.6}{rm}C9}}}
\put(226,-5311){\makebox(0,0)[lb]{\smash{\SetFigFont{8}{9.6}{rm}C6}}}
\put(901,-6736){\makebox(0,0)[lb]{\smash{\SetFigFont{8}{9.6}{rm}C8}}}
\put(226,-6736){\makebox(0,0)[lb]{\smash{\SetFigFont{8}{9.6}{rm}C4}}}
\put(901,-7336){\makebox(0,0)[lb]{\smash{\SetFigFont{8}{9.6}{rm}C3}}}
\put(526,-6061){\makebox(0,0)[lb]{\smash{\SetFigFont{8}{9.6}{rm}C2,C5,C7}}}
\put(151,-4561){\makebox(0,0)[lb]{\smash{\SetFigFont{8}{9.6}{rm}(c)}}}
\put(2401,-4561){\makebox(0,0)[lb]{\smash{\SetFigFont{8}{9.6}{rm}(d)}}}
\put(3601,-4636){\makebox(0,0)[lb]{\smash{\SetFigFont{8}{9.6}{rm}$\overline{C3}$}}}
\put(3601,-5311){\makebox(0,0)[lb]{\smash{\SetFigFont{8}{9.6}{rm}$\overline{C4}$}}}
\put(3301,-6061){\makebox(0,0)[lb]{\smash{\SetFigFont{8}{9.6}{rm}$\overline{C2}$, $\overline{C5}$}}}
\put(4501,-6061){\makebox(0,0)[lb]{\smash{\SetFigFont{8}{9.6}{rm}$\overline{C8}$}}}
\put(3601,-6811){\makebox(0,0)[lb]{\smash{\SetFigFont{8}{9.6}{rm}$\overline{C6}$}}}
\put(4501,-6811){\makebox(0,0)[lb]{\smash{\SetFigFont{8}{9.6}{rm}$\overline{C7}$}}}
\put(2251,-6886){\makebox(0,0)[lb]{\smash{\SetFigFont{8}{9.6}{rm}$\overline{C1}$, $\overline{C9}$}}}
\end{picture}
\vspace{.2in}
\caption{Comparison of overlapping cost drivers in Extrusion and Forging for $\gamma$-Titanium LPT Cover Plate}
\end{figure}
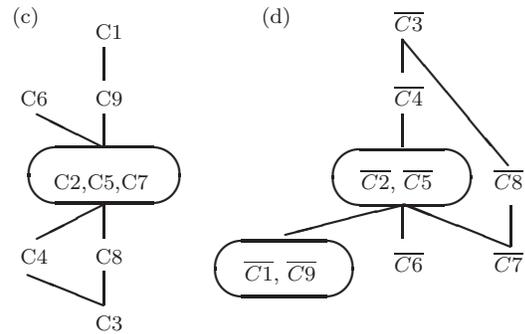

So far we have discussed some (but not all) results of analysis 
we have performed at {\sf the level of component parts of an aircraft 
jet engine}. 
We have, however, also achieved significant results in developing 
new methods for the {\sf level of integration of components into a 
subsystem} as well as integration and summarization of data 
creating the levels of coarser granularity. We shall survey some of 
these now. 
 
The Fuzzy Relational Affordability Systemic Model (FRASMod)  has been 
designed to capture and integrate diversity of contexts and perspectives 
of manufacturing activities within a unified representation structure. 

We have seen that in building of FRASMod the key entities of each perspective are 
identified using the exploratory knowledge elicitation  and mapped into 
a relational subsystem and a relational coupling structure that shows 
potential interactions of the entities corresponding to different 
perspectives. 

It is not only integration, but also information summarization 
that is essential at this level of knowledge representation. 
We have developed three techniques for this purpose. Namely, 

\noindent 
\hspace*{ 0.9 cm} $\bullet$ Interval Aggregation of costs: A fuzzy 
algorithm for computing
interval bounds of the cost of the subsystem as a function of parts and
values of the process attributes. (See Sec. C.3.4 below) \\
\noindent 
\hspace*{ 0.9 cm} $\bullet$ The method of summarization of preorders 
to provide an interval ranking of objects or attributes. \\
\noindent 
\hspace*{ 0.9 cm} $\bullet$ Generalized morphisms based comparison 
of structures: for relating information concerning 
the structures of resolution levels and correct aggregation of measurements.

\subsection{Interval Aggregation of Costs}  

Based on the possibility measure and the plinth of fuzzy sets \cite{80.1} 
we have developed an interval  method for the computing the interval 
bounds of the affordability information  
to be used for its integration and summarization when  
moving form a lower resolution level to a higher one 
in our relational knowledge representation scheme FRASMod,  
just creating the levels of coarser granularity. 

This method has been used for computing interval bounds of the aggregated 
cost that is the  function of the values of the  the 86 cost drivers of 
the five processes involved in manufacturing the LPT cover plate. 
The same procedure can be applied recursively, to yield the interval bound 
on the total cost of integrating the LPT cover plate with other parts of a
Low Pressure Turbine. Further higher recursion is also possible. So
the method applicable on any level, suitably using aggregated information
from the lower levels.

\section{Use of  preorders to provide an 
interval ranking of competing technologies}

Fig. 8 shows a demonstration example of fuzzy interval ranking of
technologies using this set of data\footnote{
This solves a problem 
proposed to us by our industrial partner Pratt \& Whitney.}.         

Given parameters of selected 
technologies, {\it Investment priority}   
partial ordering  of {\it technologies preferences} 
can be  computed. The Hasse diagrams then express the partial 
ranking of technologies based on parameters 
such as {\it potential investment, Improvement of performance} and 
various {\it potential} and {\it benefit measures}. 
The evaluated objects are technologies $T_1$ to $T_7$, that are 
characterized by seven attributes $P_1$ to $P_7$ as shown at the top of Fig.
8. The result of relational analysis are the Hasse diagrams displayed 
at the bottom of Fig. 8. 

It can be seen from the Hasse diagrams that processing the data by 
different fuzzy logics yields different partial ordering of technologies. 
Hence, the input data is fits several competing models. To reconcile the 
differences 
we have to collect more data or use interval fuzzy logics. 

This interval method that we have developed uses one of the Checklist paradigm 
\cite{96.3} based interval systems, a triple 
$<$ {\L}ukasiewicz, Reichenbach, Kleene-Dienes $>$ 
logics combined with appropriate summarization procedure. 
The initial interval ranking obtained by applying the FIRE procedures 
to the sets of 
Hasse diagrams of technologies  is displayed 
at the bottom of Fig. 8 at the right (for the $\alpha$-cut with 
value 0.17).  

In exploratory analysis of possible technological alternatives where only 
few global cost characteristics of the technologies are available, the
preference is usually expressed by linear ranking done heuristically. 
If, however, the intrinsic order contained in the data is only a partial
ordering, the linear order is usually enforced
artificially, e.g by an accountant or economist disregarding whether or 
not it is this linear order is intrinsically present in the data. 
The case of such heuristic ranking by an engineer is displayed in the 
right column of the table depicted in Fig. 8. The artificial 
unwarranted precision is introduced. Compare it with the ranking 
intervals computed by FIRE displayed in the same figure. 

Clearly, FIRE does not impose linear ranking when it is not 
present in the given data. It will come out if it is there. 
But where there is only little information, 
with large ``grey bands" of imprecision, our method does not artificially 
impose it, but works with intervals instead. 

The importance of FIRE goes beyond just ranking technologies 
stems from the fact that the {\bf FIRE method can be applied at any lower 
resolution level}: E.g. objects are not technologies but alternative 
manufacturing processes by which a component can be produced. The 
of the cost factors when integrating  components into a 
subsystem.attributes may be cost drivers, performance measures, 
reliability measures etc. 

{\small 
\vspace{-.2in}
\noindent
\begin{table} [htbp]
\caption{Original Data from Pratt \& Whitney }
      \begin{tabbing}
      $ORG = T \times P$ \\
      where \= $T = \{ T_{1},T_{2},T_{3},T_{4},T_{5},T_{6},T_{7} \}$,  a set of technologies   \\   
            \> $P = \{ P_{1},P_{2},P_{3},P_{4},P_{5},P_{6},P_{7} \}$,  a set of cost parameters
       \end{tabbing}


\hspace*{-1.1cm}
\footnotesize
 \begin{tabular}{|c||c|c|c|c|c|c|c||c|} \hline
  \multicolumn{1}{|c||}{}
& \multicolumn{7}{c||}{Parameters}
& \multicolumn{1}{c|}{}\\ \cline{2-8}
  \multicolumn{1}{|c||}{}
& \multicolumn{1}{c|}{}
& \multicolumn{1}{c|}{}
& \multicolumn{1}{c|}{}
& \multicolumn{1}{c|}{}
& \multicolumn{1}{c|}{}
& \multicolumn{1}{c|}{}
& \multicolumn{1}{c||}{Non Weighted}
& \multicolumn{1}{c|}{} \\
  \multicolumn{1}{|c||}{}
& \multicolumn{1}{c|}{}
& \multicolumn{1}{c|}{}
& \multicolumn{1}{c|}{}
& \multicolumn{1}{c|}{}
& \multicolumn{1}{c|}{}
& \multicolumn{1}{c|}{}
& \multicolumn{1}{c||}{Economic,}
& \multicolumn{1}{c|}{} \\
  \multicolumn{1}{|c||}{}
& \multicolumn{1}{c|}{Potential}
& \multicolumn{1}{c|}{Cost}
& \multicolumn{1}{c|}{Improvement}
& \multicolumn{1}{c|}{Performance}
& \multicolumn{1}{c|}{Enabling}
& \multicolumn{1}{c|}{Enabling}
& \multicolumn{1}{c||}{Performance \&}
& \multicolumn{1}{c|}{Investm.} \\
  \multicolumn{1}{|c||}{}
& \multicolumn{1}{c|}{Investment}
& \multicolumn{1}{c|}{Reducing}
& \multicolumn{1}{c|}{in}
& \multicolumn{1}{c|}{Improvement}
& \multicolumn{1}{c|}{Technology}
& \multicolumn{1}{c|}{Technology}
& \multicolumn{1}{c||}{Enabling}
& \multicolumn{1}{c|}{Priority} \\
  \multicolumn{1}{|c||}{}
& \multicolumn{1}{c|}{}
& \multicolumn{1}{c|}{Potential}
& \multicolumn{1}{c|}{Performance}
& \multicolumn{1}{c|}{Potential}
& \multicolumn{1}{c|}{Benefits}
& \multicolumn{1}{c|}{Potential}
& \multicolumn{1}{c||}{Technology}
& \multicolumn{1}{c|}{} \\
  \multicolumn{1}{|c||}{}
& \multicolumn{1}{c|}{}
& \multicolumn{1}{c|}{}
& \multicolumn{1}{c|}{}
& \multicolumn{1}{c|}{}
& \multicolumn{1}{c|}{}
& \multicolumn{1}{c|}{}
& \multicolumn{1}{c||}{Improvement}
& \multicolumn{1}{c|}{} \\
  \multicolumn{1}{|c||}{Technol.}
& \multicolumn{1}{c|}{}
& \multicolumn{1}{c|}{}
& \multicolumn{1}{c|}{}
& \multicolumn{1}{c|}{}
& \multicolumn{1}{c|}{}
& \multicolumn{1}{c|}{}
& \multicolumn{1}{c||}{Potential}
& \multicolumn{1}{c|}{} \\
& \multicolumn{1}{c|}{$P_{1}$}
& \multicolumn{1}{c|}{$P_{2}$}
& \multicolumn{1}{c|}{$P_{3}$}
& \multicolumn{1}{c|}{$P_{4}$}
& \multicolumn{1}{c|}{$P_{5}$}
& \multicolumn{1}{c|}{$P_{6}$}
& \multicolumn{1}{c||}{$P_{7}$}
&  \\ \hline
& \multicolumn{1}{c@{\,}} {$\mid$\$$\mid$L\$} & {Indicator} & {\% \& Fault} & {Indicator} & {\% \& A/C Cost} & {Indicator} & {Indicator} & {Ranking} \\ \hline \hline
$T_{1}$ & 65 & 90  & 1.00\% & 46  & 0.13\% & 56  & 64  & 4  \\ \hline
$T_{2}$ & 90 & 45  & 1.20\% & 40  & 0.10\% & 32  & 39  & 7  \\ \hline 
$T_{3}$ & 30 & 186 & 1.20\% & 120 & 0.27\% & 263 & 190 & 2  \\ \hline  
$T_{4}$ & 60 &  3  & 2.00\% & 100 & 0.28\% & 124 & 75  & 3  \\ \hline  
$T_{5}$ & 25 & 14  & 0.80\% & 98  & 0.04\% & 45  & 52  & 6  \\ \hline  
$T_{6}$ & 100 & 14 & 1.80\% & 48  & 0.03\% &  9  & 24  & 8  \\ \hline  
$T_{7}$ & 75 &  5  & 2.00\% & 80  & 0.21\% & 80  & 55  & 5  \\ \hline  
\end{tabular}

\end{table}
}

{\footnotesize
\begin{figure}[htbp]
\caption{Interval Ranks of Technologies and their HD structures}
\vspace{.2in}
\hspace*{-1cm}
\setlength{\unitlength}{0.010000in}%
\begingroup\makeatletter
\def\x#1#2#3#4#5#6#7\relax{\def\x{#1#2#3#4#5#6}}%
\expandafter\x\fmtname xxxxxx\relax \def\y{splain}%
\ifx\x\y   
\gdef\SetFigFont#1#2#3{%
  \ifnum #1<17\tiny\else \ifnum #1<20\small\else
  \ifnum #1<24\normalsize\else \ifnum #1<29\large\else
  \ifnum #1<34\Large\else \ifnum #1<41\LARGE\else
     \huge\fi\fi\fi\fi\fi\fi
  \csname #3\endcsname}%
\else
\gdef\SetFigFont#1#2#3{\begingroup
  \count@#1\relax \ifnum 25<\count@\count@25\fi
  \def\x{\endgroup\@setsize\SetFigFont{#2pt}}%
  \expandafter\x
    \csname \romannumeral\the\count@ pt\expandafter\endcsname
    \csname @\romannumeral\the\count@ pt\endcsname
  \csname #3\endcsname}%
\fi
\endgroup
\begin{picture}(535,190)(-105,640)
\thicklines

\put(388,752){\circle*{6}}
\put(448,752){\circle*{6}}
\put(388,732){\circle*{6}}
\put(478,732){\circle*{6}}
\put(358,712){\circle*{6}}
\put(358,692){\circle*{6}}
\put(388,692){\circle*{6}}
\put(388,672){\circle*{6}}
\put(448,672){\circle*{6}}
\put(508,672){\circle*{6}}
\put(388,652){\circle*{6}}
\put(478,652){\circle*{6}}
\put(358,632){\circle*{6}}
\put(418,632){\circle*{6}}

\put(388,752){\line( 1, 0){ 60}}
\put(388,732){\line( 1, 0){ 90}}
\put(358,692){\line( 1, 0){ 30}}
\put(448,672){\line( 1, 0){ 60}}
\put(388,652){\line( 1, 0){ 90}}
\put(358,632){\line( 1, 0){ 60}}

\put(335,610){\line( 1, 0){195}}
\put(335,765){\line( 0,-1){155}}
\put(360,612){\line( 0,-1){  5}}
\put(390,612){\line( 0,-1){  5}}
\put(420,612){\line( 0,-1){  5}}
\put(450,612){\line( 0,-1){  5}}
\put(480,612){\line( 0,-1){  5}}
\put(510,612){\line( 0,-1){  5}}
\put(333,630){\line( 1, 0){  5}}
\put(333,650){\line( 1, 0){  5}}
\put(333,670){\line( 1, 0){  5}}
\put(333,690){\line( 1, 0){  5}}
\put(333,710){\line( 1, 0){  5}}
\put(333,730){\line( 1, 0){  5}}
\put(333,750){\line( 1, 0){  5}}
\put(358,692){\line( 1, 0){ 30}}

\put(440,800){\makebox(0,0)[lb]{\smash{\SetFigFont{10}{12.0}{bf}$\alpha = .17$}}}
\put(545,605){\makebox(0,0)[lb]{\smash{\SetFigFont{10}{12.0}{rm}Rank}}}
\put(315,625){\makebox(0,0)[lb]{\smash{\SetFigFont{10}{12.0}{rm}$T_{7}$}}}
\put(315,645){\makebox(0,0)[lb]{\smash{\SetFigFont{10}{12.0}{rm}$T_{6}$}}}
\put(315,665){\makebox(0,0)[lb]{\smash{\SetFigFont{10}{12.0}{rm}$T_{5}$}}}
\put(315,685){\makebox(0,0)[lb]{\smash{\SetFigFont{10}{12.0}{rm}$T_{4}$}}}
\put(315,705){\makebox(0,0)[lb]{\smash{\SetFigFont{10}{12.0}{rm}$T_{3}$}}}
\put(315,725){\makebox(0,0)[lb]{\smash{\SetFigFont{10}{12.0}{rm}$T_{2}$}}}
\put(315,745){\makebox(0,0)[lb]{\smash{\SetFigFont{10}{12.0}{rm}$T_{1}$}}}
\put(310,770){\makebox(0,0)[lb]{\smash{\SetFigFont{10}{12.0}{rm}Technology}}}
\put(360,590){\makebox(0,0)[lb]{\smash{\SetFigFont{10}{12.0}{rm}1}}}
\put(390,590){\makebox(0,0)[lb]{\smash{\SetFigFont{10}{12.0}{rm}2}}}
\put(420,590){\makebox(0,0)[lb]{\smash{\SetFigFont{10}{12.0}{rm}3}}}
\put(450,590){\makebox(0,0)[lb]{\smash{\SetFigFont{10}{12.0}{rm}4}}}
\put(480,590){\makebox(0,0)[lb]{\smash{\SetFigFont{10}{12.0}{rm}5}}}
\put(510,590){\makebox(0,0)[lb]{\smash{\SetFigFont{10}{12.0}{rm}6}}}


\put(70,623){\oval(56,24)}
\put(225,658){\oval(56,24)}
\put(-73,685){\oval(70,30)}
\put(-73,728){\oval(54,24)}

\put(175,785){\line( 1, 0){105}}
\put(225,645){\line( 0,-1){ 15}}
\put(225,685){\line( 0,-1){ 15}}
\put(225,720){\line( 0,-1){ 15}}
\put(225,755){\line( 0,-1){ 15}}
\put(225,755){\line(-1,-2){ 62}}

\put(15,785){\line( 1, 0){130}}
\put(70,755){\line( 0,-1){ 15}}
\put(70,720){\line( 0,-1){ 15}}
\put(70,685){\line( 0,-1){ 15}}
\put(70,650){\line( 0,-1){ 15}}
\put(70,610){\line( 0,-1){ 15}}

\put(-115,785){\line( 1, 0){ 95}}
\put(-73,670){\line( 0,-1){ 15}}
\put(-73,715){\line( 0,-1){ 15}}
\put(-73,755){\line( 0,-1){ 15}}


\put(-110,825){\makebox(0,0)[lb]{\smash{\SetFigFont{10}{12.0}{bf}Technology : (Operator, $\alpha$-cut, Criteria)}}}
\put(-115,820){\line( 1, 0){290}}

\put(-110,790){\makebox(0,0)[lb]{\smash{\SetFigFont{10}{12.0}{bf}({\L}, M, mean)}}}
\put(-77,640){\makebox(0,0)[lb]{\smash{\SetFigFont{10}{12.0}{rm}t5}}}
\put(-95,680){\makebox(0,0)[lb]{\smash{\SetFigFont{10}{12.0}{rm}t1, t2, t6}}}
\put(-87,722){\makebox(0,0)[lb]{\smash{\SetFigFont{10}{12.0}{rm}t4, t7}}}
\put(-77,760){\makebox(0,0)[lb]{\smash{\SetFigFont{10}{12.0}{rm}t3}}}

\put(170,790){\makebox(0,0)[lb]{\smash{\SetFigFont{10}{12.0}{bf}(KD, M, mean)}}}
\put(155,615){\makebox(0,0)[lb]{\smash{\SetFigFont{10}{12.0}{rm}t1}}}
\put(220,615){\makebox(0,0)[lb]{\smash{\SetFigFont{10}{12.0}{rm}t5}}}
\put(210,654){\makebox(0,0)[lb]{\smash{\SetFigFont{10}{12.0}{rm}t2, t6}}}
\put(220,690){\makebox(0,0)[lb]{\smash{\SetFigFont{10}{12.0}{rm}t7}}}
\put(220,725){\makebox(0,0)[lb]{\smash{\SetFigFont{10}{12.0}{rm}t4}}}
\put(220,760){\makebox(0,0)[lb]{\smash{\SetFigFont{10}{12.0}{rm}t3}}}

\put(20,790){\makebox(0,0)[lb]{\smash{\SetFigFont{10}{12.0}{bf}(KD{\L}, M, mean)}}}
\put(65,760){\makebox(0,0)[lb]{\smash{\SetFigFont{10}{12.0}{rm}t3}}}
\put(65,725){\makebox(0,0)[lb]{\smash{\SetFigFont{10}{12.0}{rm}t4}}}
\put(65,690){\makebox(0,0)[lb]{\smash{\SetFigFont{10}{12.0}{rm}t7}}}
\put(65,655){\makebox(0,0)[lb]{\smash{\SetFigFont{10}{12.0}{rm}t1}}}
\put(55,619){\makebox(0,0)[lb]{\smash{\SetFigFont{10}{12.0}{rm}t2, t6}}}
\put(65,580){\makebox(0,0)[lb]{\smash{\SetFigFont{10}{12.0}{rm}t5}}}


\put(-50,560){\makebox(0,0)[lb]{\smash{\SetFigFont{10}{12.0}{rm}t1 : technology-1}}}
\put(-50,542){\makebox(0,0)[lb]{\smash{\SetFigFont{10}{12.0}{rm}t2 : technology-2}}}
\put(-50,524){\makebox(0,0)[lb]{\smash{\SetFigFont{10}{12.0}{rm}t3 : technology-3}}}
\put(100,560){\makebox(0,0)[lb]{\smash{\SetFigFont{10}{12.0}{rm}t4 : technology-4}}}
\put(100,542){\makebox(0,0)[lb]{\smash{\SetFigFont{10}{12.0}{rm}t5 : technology-5}}}
\put(100,524){\makebox(0,0)[lb]{\smash{\SetFigFont{10}{12.0}{rm}t6 : technology-6}}}
\put(250,560){\makebox(0,0)[lb]{\smash{\SetFigFont{10}{12.0}{rm}t7 : technology-7}}}
\put(400,560){\makebox(0,0)[lb]{\smash{\SetFigFont{10}{12.0}{rm}{\L}   :  {\L}ukasiewicz}}}
\put(400,542){\makebox(0,0)[lb]{\smash{\SetFigFont{10}{12.0}{rm}KD{\L} : Richenbach}}}
\put(400,524){\makebox(0,0)[lb]{\smash{\SetFigFont{10}{12.0}{rm}KD  : Kleene-Dienes}}}
\put(250,524){\makebox(0,0)[lb]{\smash{\SetFigFont{10}{12.0}{rm}M  :  Mean $\alpha$-cut}}}
\end{picture}
\end{figure}
}

\newpage

\section{Value Analysis as a Tool for Identification of Unnecessary Costs}

 Value analysis 
[R12],\cite{zenz.bk} is the organized, systematic study of 
the function of a material, part, component, or system to identify 
areas of unnecessary cost used in any production or service.
Value analysis (VA) consists of (1) analyzing the function of a product,  
(2) considering designs to accomplish this function, and (3) analyzing 
the costs of alternatives. Activity structures methodology unlike some 
other methods can combine analysis by activities with
analysis by functions. This is made possible because it distinguishes
substratum structures, system activity structures and functional activity
structures \cite{89.5},\cite{89.1}.  

We have integrated  the Value 
Analysis \cite{zenz.bk} and  Activity Structures \cite{89.4},\cite{89.5} 
methodologies, and investigated the ways of 
building relational models for processing data generated by Value Analysis.

Relational model of Value Analysis activities 
using BK-relational products has been formulated [R12]. 
Such a model allows us to identify the {\it crucial technological 
and business factors} that influence the value of products 
and services in order to provide alternatives of better value. 
It also helps to integrate business factors with engineering 
factors and analyze these by relational computations for 
their similarity, equivalence and 
mutual dependence as described in objective 1 above.     
Currently, relational value analysis 
of the $\gamma$-titanium LPT cover plate is in progress, to 
supplement its engineering analysis by  
analysis of non-engineering factors influencing its cost. 
 
The relational model [R12] has also been used to develop a fuzzy 
algorithm for cost generalized optimization [R15]. It makes it possible to 
optimize the cost of design of a system by choosing the best alternative with
respect to cost, performance and undesirable side-effects.  \\

Value analysis is an important method for reducing the cost of manufactured
products. It is the organized, systematic study of 
the function of a material, part, component, or system to identify 
areas of unnecessary cost used in any production or service.
Value analysis (VA) consists of (1) analyzing the function of a product,  
(2) considering designs to accomplish this function, and (3) analyzing 
the costs of alternatives.

VA allows us to identify the {\it crucial technological 
and business factors} that influence the value of products 
and services in order to provide alternatives of better value. 
It also helps to integrate business factors with engineering 
factors and analyze these by relational computations for 
their similarity, equivalence and 
mutual dependence. 

 Any formal model to be practically usable has to capture the great 
diversity of factors that influence the quality of the industrial product. 
The information and data for such an analysis model are drawn  from a 
multiplicity of sources belonging to various company sections and personnel 
of different specialization. Typically, {\it ``the purpose of Value 
Analysis is to bring together ... the combined talents of purchasing 
and its vendors as well as engineering, production, and other operating 
personnel to review the components and materials used by the organization 
on products or processes already in place.  It is intended to provide 
a means of considering all possible alternatives in an atmosphere of 
open thinking and analysis."} \cite{zenz.bk}, p. 469.

Relational representation of Value Analysis data together with 
the compositions  provided by triangle and square BK-products and 
further operations over these (such as fuzzy relational closures and 
interiors) can capture the great diversity of factors, investigate their 
similarity, equivalence and mutual dependence. This helps in 
identifying the crucial factors that influence the value of products 
and services and also in providing alternatives of better value.

Different sources and different knowledge domains entering 
the overall purchasing, design, manufacturing and marketing 
activities are in reality mixed together in a multiplicity of 
contexts. The modeling apparatus on which we base our computer
support of Value Analysis must also  possess the capability of 
dealing with a number of diverse contexts \cite{89.1}. 
In each domain, appropriate contexts must be distinguished. 
In setting a relational model, one has to clearly understand what 
is the meaning of the key notions in individual contexts, 
and what role these  play. 
In Value Analysis one wants to find regular phenomena and intrinsic 
dependencies of various factors within complex interrelationships 
of all factors and contexts into which the product enters. 
Within this framework we are specifically interested in detection of 
change to identify those trends that need to be encouraged or curtailed 
as the case may be. Interdependencies of various factors, parts, 
subsystems and observables characterizing the evaluated product  and their 
links with cost and utility have to be established. 

The following name-lists(see Fig.9) specify the concepts used in one of our VA relational
models developed in this project. (In Knowledge Engineering these are called
lists of 'ontologies' or sometimes `semiotic descriptors').

\begin{figure}

{\small 
\begin{tabbing}
bla \= bla \= blahblahblahblahblahblahblah \kill
{\sf Name:} \> \> {\sf A set of:} \\[.6em]
B \> .... \> Systems of functions. \\
C \> .... \> Cost.  \\                      
G \> .... \> Processes.\\ 
H \> .... \> Substratum units (physically related subsystems of components, etc.)\\
I \> .... \> Investigations (quality tests, etc.). \\    
M \> .... \> Modifications.\\
O \> .... \> Observation events (e.g. time indexing). \\
P \> .... \> Part or Component.\\
S \> .... \> Observable features, measurable properties, functional signs.  \\           
U \> .... \> Usability measure. \\  
V \> .... \> Variant of a substratum unit/module (e.g. a part). \\
Y \> .... \> Composed attributes, functional characteristics.
\end{tabbing} 
}

\noindent
A number of meaningful relations between these entities can be 
formed. 

{\small 
\begin{tabbing}
blahbl \= blalbb \= blahblahblahb \= blahblahblahblahblahblahblah \kill
{\sf Name/Type:} \> \> {\sf Definition:} \> {\sf Relation} \\[1em]
blah \= bla \= blahblahblahbla \= blahblahblahblahblahblahblah \kill
VYC \> .... \> ${\cal R}(V \times Y \times C)$   
\> between {\it variants of a part, functional features} and {\it cost}. \\
VYU \> .... \> ${\cal R}(V \times Y \times U)$   
\> between {\it variants of a part, functional features} and {\it usability}. \\
BYS \> .... \> ${\cal R}(B \times Y \times S)$   
\> between {\it systems of functions, functional characteristics} and  {\it properties}. \\
PVC \> .... \> ${\cal P}(P \times V \times C)$   
\> between {\it parts, variants of parts} and {\it cost}. \\
PVY \> .... \> ${\cal P}(P \times V \times Y)$   
\> between {\it parts, variants of parts} and {\it functional features}. \\
PVU \> .... \> ${\cal P}(P \times V \times U)$   
\> between {\it parts, variants of parts} and {\it usability}. \\
 PY \> .... \> ${\cal R}(P$ {\normalsize $\leadsto$} $Y)$ 
\> from {\it parts} to {\it functional features} \\[-1em]
\end{tabbing} 
}
\caption{The conceptual meaning of sets and relations used in the value 
analysis example below}
\end{figure}

The following examples show the use of relational models. 
For the explanation of the notation used, see Appendix 1.
$\bigoplus_{j}$ is an aggregation operator; its simple instance is
e.g. $\displaystyle{\frac{1}{n} \sum^{n}_{j=1}}$.

\noindent
{\bf Example of relational computations in value analysis:} 
We wish to compare different parts with respect to their
functional features using the entities listed in Fig.9.

\noindent
Let PY be a relation from the set of parts $P$ to the set $Y$ of functional 
features. The triangle subproduct \[(PY \lhd PY^{T})_{ik} =  
\bigoplus_{j} (PY_{ij} \equiv PY^{T}_{jk}) \]
will give  {\it the degree to which the functional features of
part $p_{i}$ are included in the set of functional features of part
$p_{k}$.}

\noindent
Let PYC be a ternary (3-place) relation between the set of parts $P$, 
the set $Y$ of functional features and the set $C$ of costs. The square 
product \[(PYC \Box PYC)_{ijlm} =  
\bigoplus_{k=n} (PYC_{ijk} \equiv PYC_{lmn}) \]
will give {\it the degree to which the cost of variant $v_i$ 
of part $p_i$ matches the cost of variant $v_m$ of part $p_l$ }.

\noindent
Let VYC be a ternary (3-place) relation between the set of variants of a
part,  the set $Y$ of functional features and the set $C$ of costs. 
The square 
product \[(VYC \Box VYC)_{ikln} =  
\bigoplus_{j=m} (VYC_{ijk} \equiv VYC_{lmn}) \]
will give  {\it the degree to which variant $v_i$ 
of a part costing $c_k$ is exchangeable for  variant $v_l$ 
of the same part with respect to matching their functional features.} 

Many other relevant VA-questions can answered be by various combinations of
fuzzy BK-products and the answers can be ranked by the degree of validity or
relevance of the answer within a specific context. 

The context addressed in this project that are of particular interest to our
industrial partner Pratt\&Whitney are depicted in the Figure 1 above, in Sec.1.3.

\section{Appendix 1: A Survey of Theory and Applications of
Fuzzy BK-Products}

\subsection{The Unifying Power of Relations} 

Relational representation of knowledge
 makes it possible to perform all the computations and 
decision making in a uniform relational way \cite{92.7}, by means 
of {\it special relational compositions} called triangle and square 
products. These were first introduced by Bandler and Kohout in 1977 
\cite{87.8},\cite{80.3},\cite{77.rel}
and are referred to as the BK-products  in the literature 
\cite{hajek.bk},\cite{deb+ker93pr},\cite{deb+kerMP}. Their 
theory and applications have made substantial progress since then.  

Triangle relational products together with fast fuzzy 
relational algorithms \cite{82.3},\cite{88.3}
have been applied to various practical problems in a number of 
scientific fields:  computer protection and AI \cite{89.5}, 
medicine, information retrieval,
handwriting classification, 
architecture and urban studies,  investment 
and control fields \cite{89.1}.  
See the survey in \cite{92.7} with a list of 
50 selected references on the theory and applications 
The relational methods combining linguistic labels with BK-products 
give a natural conceptual framework for knowledge 
representation and inference from imprecise, incomplete, or not totally 
reliable information in a consistent manner.
All these approaches may be enriched by {\it extending} these to the 
realm of {\bf interval computations}. 
For example or knowledge-based medical system {\sc Clinaid} combines fuzzy
relations, with methods of interval inference \cite{89.1}.

There are several 
types of product used to produce product-relations \cite{87.8}, 
\cite{92.7},\cite{80.2}. 
\begin{defin}  
For arbitrary fuzzy relations in [0, 1], $R$ from the set $X$ to $Y$, $S$
from $Y$ to $Z$ define: \\
1. \( R \circ S = (\forall x) (\forall z) (\exists y) (xRy \; \& \; ySz); \) 
\hspace{.5in} 2. \( R \lhd S  =(\forall x) (\forall z) (\forall y) (xRy \rightarrow ySz);
\) \\ 
3. \( R \rhd S  = (\forall x) (\forall z) (\forall y) (xRy \leftarrow  ySz);
\) \hspace{.5in} 4. \( R \Box S  = (\forall x) (\forall z) (\forall y) 
(xRy \equiv  ySz)\) \\[- 0.6cm] 
\end{defin}

\noindent 
Only the conventional $\circ$ 
is associative. The triangle and square products, on the other hand,  have
important properties that give the 
power and versatility to our methods of relational analysis. 
$\Box$is not associative at all, and the 
the following pseudo-associativities hold: \cite{86.2}: \\
1. $Q \lhd (R \rhd S) = (Q \lhd R) \rhd S$, \hspace{.1in}
2. $Q \lhd (R \lhd S) = (Q \circ R) \lhd S$, \hspace{.1in}
3. $Q \rhd (R \rhd S) = Q \rhd ( R \circ S)$.

On the abstract side of non-fuzzy (crisp) relational algebras (RA), Tarski
and his school  have investigated the interrelationship of various 
RAs. Namely, representable (RRA), semiassociative (SA), weakly 
associative (WA) and non-associative (NA) relational algebras. Maddux 
\cite{maddux.varra} gives the following result: \\ 
RRA $\subset$ RA $\subset$ SA $\subset$ WA $\subset$ NA. These results do
not say anything about representations of these extended relational
algebras.  

The  BK-products defined over relational calculi give the constructive
realization of the non-associative products for both crisp and fuzzy
relations. Hence, non-associative products
have representations and that these products offer various computational 
advantages. For example, the following {\bf universal representation of
preorders} is given for  all the  relations that are in the lattice ${\cal
R}(X \leadsto X)$:   

\begin{theor} \cite{86.2}  
(a) $R$ is a preorder if and only if  $R = R \rhd R^{-1}$.      \\
(b) Every preorder or relations can be expressed that way. \\
(c) $R= R \Box R^{-1}$ if and only if $R$ is an equivalence.
\end{theor}

\noindent 
 $\lhd, \rhd, \Box$ products add the expressive power to the 
mathematics of relations. Very important for distributed knowledge
networking  is a constructive generalization of conventional homomorphisms
defined constructively by BK-products:  

\begin{defin}
Let $F, R, G, S$ be the relations between the sets $A, B, C, D$ 
such that $R \in {\cal R}(A \leadsto B)$. The conditions that 
(for all $a \in A, b \in B, c \in C, d \in D$) 
$aFc$ and $aRb$ and $bGd$ imply $cSd$, will be expressed in any of 
the following ways: 
(i) $FRG; S$ are {\bf forward compatible}  
(ii) $F, G$ are {\bf generalized homomorphisms} from $R$ to $S$.
\end{defin} 

\begin{theor} {\it Compatibility} \cite{86.2}) \textsf{ }
1. $FRG; S$ are forward compatible if and only if $F^{T} \circ R 
\circ G \sqsubseteq S$.\\
2. {\it Formulas for computing the explicit compatibility 
criteria for F and G} are: \                   
$FRG; S$ are forward-compatible iff 
$F \sqsubseteq R \lhd (G \lhd S^{T})$ 
\end{theor}

Similarly, the {\it backward compatibility} is defined and constructive 
conditions for relations both-way compatible   (i.e. forward 
{\it and} backward) given \cite{86.2}. {\it Both-ways} compatibility subsumes 
the conventional homomorphisms.

\subsection{Dealing With Incomplete and Uncertain Information}

 Expert reasoning, decision making and actions have to operate 
on the background of uncertainty, incompleteness of information 
and conflicting evidence. These activities involve conceptual 
structures and dispositions that the experts intuitively use. It 
also involves reference to linguistic structures and their 
capability to handle multiple contexts. Understanding these 
underlying processes is difficult, yet essential in our attempts 
to aid expert decision making with computing and information 
processing technology.

 Reasoning with uncertainty, incompleteness and also with 
conflicting evidence (to be called {\it reasoning with imperfect 
information}) cannot be fully devoid of the conceptual 
structures upon which the phenomena of vagueness, uncertainty, 
incompleteness of information and conflicting evidence operate. 
Identification of relevant conceptual structures, 
meta-frameworks, frameworks and knowledge contents of individual 
knowledge domains therefore plays the crucial role in such 
reasoning with imperfect information.

 It is not only the syntactic structure and logical form , but 
also the complete linguistic structure, including the semantic 
contents and other semiotic aspects that is important. For this 
reason, even partial attempts at capturing the essential 
features of expert's competence in our information processing 
technology require new tools and new architectures that are 
capable of dealing fully with these aspects. Otherwise, the 
richness of the conceptual and linguistic world of a competent 
expert would be distorted beyond recognition, with side effects 
on our everyday life that can be disastrous. Thus we have to 
face the problem of systematizing and formalizing the semantics 
of expert actions acquisition, representation and utilization of 
knowledge in a new way. This approach has to have special 
features: it is to be generally {\bf context-dependent} where 
{\bf localized} relevant fragments of knowledge, form a system; 
and within this system, reasoning with imperfect information 
ought to operate adequately. 

 Such a unification requires a formal descriptive and 
computational approach that would put on equal footing the 
conceptual, linguistic and semiotic part with the mathematical 
computational part. One has also face the problem of conceptual 
conflicts \cite{man.kbs8} and of their resolution. This may 
leads directly to paraconsistent logics \cite{91.17}. 
This unification can be achieved by relational method using
BK-products of relations.

\subsection{A Brief Overview of Fuzzy BK-Products}

\noindent 
{\bf Mathematical definitions.} \ \  Where $R$ is a relation 
from $X$ to $Y$, and $S$ a relation 
from $Y$ to $Z$, a {\it product relation} $R*S$ is a relation 
from $X$ to $Z$, determined by $R$ and $S$. There are several 
types of product used to produce product-relations \cite{87.8}, 
\cite{92.7}. Each product type performs a {\bf different logical 
action} on the intermediate sets, as {\it each logical type} of 
the product enforces a {\it distinct specific meaning} on the 
resulting product-relation $R*S$.  We have the following definitions 
of the products.  In these definitions, $R_{ij}, S_{jk}$ represent the 
fuzzy degrees to which the respective statements $x_{i}Ry_{j}$, 
$y_{j}S_{jk}z_{k}$ are true.

\begin{tabbing}

blahblahblahblahblahblahblah \= blahblahblahblahblahblahblahblah
\=   blahblahblahblahblahblah  \kill

{\sc Product Type} \> {\sc Set-based Definition} \> {\sc Many-Valued Logic
Formula}\\

{\bf Circle} product: \> $ x(R \circ S)z \Leftrightarrow xR$
intersects $Sz$ \> $(R \circ S)_{ik}= {\bigvee}_{j} (R_{ij}
\bigwedge S_{jk})$ \\        

{\bf Triangle Subproduct}: \>
$ x(R \lhd S)z \Leftrightarrow xR \stackrel{\sim}{\subseteq} Sz$ \>
$(R S)_{ik}= {\bigwedge}_{j} (R_{ij} \leadsto S_{jk})$ \\

{\bf Triangle Superproduct}: \>
$ x(R \rhd S)z \Leftrightarrow xR \stackrel{\sim}{\supseteq} Sz$ \>
$(R \rhd S)_{ik}= {\bigwedge}_{j} (R_{ij} \leftarrow S_{jk})$ \\

{\bf Square} product: \>
$ x(R \Box S)z \ \Leftrightarrow xR \cong Sz$ \>
$(R \Box S)_{ik}= {\bigwedge}_{j} (R_{ij} \equiv S_{jk})$

\end{tabbing}

\noindent The table of definitions 
given above contains two different notational forms: 
(1) The notation using the concept of set inclusion 
and  equality  \cite{80.1},\cite{80.3}.
(2) Many-valued logic based notation, which uses the logic 
connectives $\bigwedge$ and  $\leftarrow$.  
These two different forms of relational compositions are 
algebraically equivalent, producing the same mathematical results. 
Distinguishing these forms is, however, important when constructing 
fast and efficient computational algorithms. 

The logical symbols for the logic connectives {\it AND, OR},
both {\it implications} and the {\it equivalence} in the above
formulas represent the connectives of some many-valued
logic, {\bf chosen} according to the properties of the products
required. {\it Harsh} fuzzy products (defined above) are distinguished
from the family of {\it mean} products. Given the general formula
$(R@S)_{ik} ::=\ \# (R_{ij}*S_{jk})$, \ a mean product is obtained by
replacing the outer connective \# by $\sum$ and normalizing the
resulting product appropriately. The details of choice of the
appropriate many-valued connectives are discussed in \cite{80.2},
\cite{86.5},\cite{87.6},\cite{87.13},\cite{92.7}.

\subsection{From Abstract Relations to Conceptual Meaning of Fuzzy 
Relational Structures}

\noindent 
To have abstract relations is not enough. Each relations must  
possess a clearly defined meaning  giving it a concrete practical 
linguistic interpretation within the domain of its application. 
This interpretation is provided by means of interpretable {\it linguistic
labels} of special kind that will be called {\bf semiotic descriptors}. 
The difference between ordinary linguistic label and a semiotic descriptor is
that the latter kind is subject to some  constraints determined by the
ontology of the specific domain of engineering, science or business practices.  
The assignment of semiotic descriptors
also partially determines the  linguistic meaning of the composed relation
computed by the relational  product.

A simple, but useful general relational model relates semiotic descriptors 
of two kinds:  {\it objects} and {\it properties}. To provide a 
semantic interpretation of the relations involved, we have to select the 
appropriate concepts from the domain of our interest as the names of the 
sets that enter into a relationship. Let us look at a simple example 
from  the medical domain using concepts everyone is familiar with.  
The objects can be concrete (e.g. patients) or 
abstract (diseases), the properties of these objects 
being signs, symptoms or clinical test results or constructs of  
some clinical psychological tests.

If $R$ is the relation between
{\it patients} and {\it individual symptoms}, and $S$ a relation
between {\it symptoms} and {\it diseases}, $R*S$ will be a relation
between {\it patients and diseases}. The diagnostic clinical interpretation
of each distinct logical type (e.g. the triangular square product types)
of these product-relations has a {\bf distinct clinical meaning}:\\
\noindent $ x(R \circ S)z$
$: \; \;$ degree to which patient $x$ has at least one
symptom of illness $z$.\\
$ x(R \lhd S)z$$: \; \;$ degree to which $x$'s symptoms are
among those which characterize $z$.\\
$ x(R \rhd S)z$$: \; \;$ degree to which $x$'s symptoms include
all those which characterize $z$.\\
$ x(R \Box S)z$$: \; \;$
degree to which $x$'s symptoms are exactly those of illness
$z$.

\subsection{Comparison of Structures and Investigating Their 
Properties}

 BK-relational product can be used to compare relational 
 structures. 
 Thus, if $R$ is any relation
 (perhaps itself a product of other relations) from $X$ to 
 $Y$ 
 ${\cal R}(X \leadsto Y)$ and $R^{T}$ its {\it transpose}, 
 then the product $R*R^{T} \in {\cal R}(X \leadsto X)$ 
 (where $* \in \{\circ, \lhd, \rhd, \Box\}$) might exhibit some 
 relational properties that reveal important characteristics
 of the source of information from which
 they were generated. 

Here is an example still from the medical fields using the specific 
relations mentioned above:

\noindent
$x_{i}(R \lhd R^{T})x_{k}$: \ \  patient $x_{i}$'s symptoms are among these
of $x_{k}$

\noindent
$x_{i}(R \Box R^{T})x_{k}$: \ \  patient $x_{i}$ has exactly the same
symptoms as $x_{k}$

\noindent
$y_{j}(R^{T} \lhd R)y_{l}$: \ \  whenever symptom $y_{j}$ occurs, so does
$y_{l}$ (in this group of patients)

\noindent
$y_{j}(S \Box S^{T})y_{l}$: \ \  symptom $y_{j}$ characterizes exactly
exactly the same diseases as does $y_{l}$

        Relations so constructed might exhibit some important
relational properties that reveal important characteristics
and interrelationships of the source of information from which
they were generated. Hence, methods for detecting various relational 
properties of given relations are important.

Relational properties, such as 
 reflexivity, 
 symmetry, and transitivity, and classes such as tolerances, 
 equivalences and partial orders can be extracted from the linguistic 
 information elicited by repertory grids.

 Closures and interiors of relations 
 \cite{88.3},\cite{87.8} 
 play an important role in design of fast fuzzy relational 
 algorithms 
 used in our approach. The idea of {\it comparison of a 
 relation with 
 its closure} and 
 {\it comparison of a relation with its interior} leads to 
 design and to 
 validity proofs of fast fuzzy relational algorithms (FFRA) 
 that can test 
 various local properties and also automatically discover the 
 cases 
 when the tested properties hold not only locally, but also 
 globally.

In the general terms, the abstract 
theoretical tools supporting identification and representation of 
relational properties are fuzzy closures and interiors 
\cite{88.3},\cite{82.3}. Having 
such means for testing relational properties opens the avenue to 
linking the empirical structures that can be observed and captured 
by fuzzy relations with their abstract, symbolic representations 
that have well defined mathematical properties.

Standard relational properties (both crisp and fuzzy), such as reflexivity, 
symmetry, and  transitivity, and classes such as tolerances, 
equivalences and partial orders are well understood. 
One essential drawback that both the crisp (non-fuzzy) and standard fuzzy 
theories of relational properties share is that they are defined as global, 
i.e. the properties must be be shared by all the elements of a relation.
The contributions of Bandler and Kohout crucial for multi-level 
knowledge representation investigated in this project was to 
provide an adequate {\bf definition of locality} for both crisp 
(non-fuzzy) and fuzzy relations \cite{87.8},\cite{88.3}) 
and develop software tools for computational testing of local properties 
and comparing partial relational structures.

\subsection{Multidisciplinary Work}

BK-relational products and  fast fuzzy 
relational algorithms \cite{82.3}, \cite{88.3}
were applied in numerous multi-disciplinary application:  
medical  AI, \cite{79.1}, \cite{80.2},\cite{84.2}, \cite{86.5},  
 \cite{89.7}, \cite{87.5}; information 
retrieval \cite{82.2}, \cite{83.1}, \cite{84.1}, \cite{87.4}, 
handwriting classification \cite{86.9}, natural language 
understanding \cite{91.2},\cite{91.4},\cite{nagar.phd},  generating 
efficient search strategies for resolution-based theorem 
proving \cite{92.15}, \cite{92.1}, cognitive structure analysis 
and other areas \cite{85.3}, \cite{86.18}, \cite{83.2}. A very 
promising recent application is concerned with generating 
efficient search strategies for resolution-based theorem 
proving \cite{92.15}, \cite{92.1} and in engineering and manufacturing 
 \cite{95.17}, \cite{96.5},\cite{96.6}, \cite{96.3}, \cite{96.17}, \cite{96.18}, \cite{96.11},  \cite{97.1},\cite{97.2}, \cite{97.3}, \cite{97.14}, \cite{97.6}, \cite{97.9},\cite{noe.phd}, \cite{97.12}, \cite{97.15}, \cite{97.11}, \cite{97.13} 

Relational computations are inherently parallel, hence well suited for
developing data analysis, design and decision making tools on distributed 
networked systems. This is a feature necessary e.g. for distributed
manufacturing.

\section{Appendix 2: Publications resulting from  DMI 952 5991 Project} 

\noindent           
[R1] \  P.~H{\'{a}}jek and L.J. Kohout. \ \ Fuzzy implications and 
generalized quantifiers.
{\em Internat. Journal of Uncertainty, Fuzziness and Knowledge Based
  Systems}, 4(3):225--233, 1996. [Also partially supported by 
the National Research Council COBASE grant for 
collaborative projects with former East European countries.]\\

\noindent 
[R2] \  L.J. Kohout. \ \ An invited lecture on 
{\it Fuzzy Sets in Data Analysis}.\\
Joint Statistical Conferences, American Statistical Association. 
August 1996. (30 minutes)\\

\noindent 
[R3] \  L.J. Kohout and E.~Kim. \ \ Relational algorithms for 
theoretical and empirical evaluation of
  systems of fuzzy connectives.
 In D.~Kraft, editor, {\em Proc. of IEEE Internat. Conf. on Fuzzy
  Systems}, IEEE, New York, 1996, vol. 2, pp. 918-923.\\

\noindent 
[R4] \  E. Kim, L.J. Kohout,  B.~DuBrosky and W. Bandler.\ \  Use of 
fuzzy relations for affordability decisions in high technology.
 In {\em Applications of Artificial Intelligence in Engineering XI.} 
R.A. Addey, G. Rzevski and A.K. Sunol (eds.). Computational Mechanics
 Publications, Boston 1996.  \\

\noindent 
[R5] \  L.J. Kohout. \  \  Decision-Making with Incomplete 
Information in an Integrated 
Product and Process Development Enterprize -- A Management 
Decision Tool for Cost Modeling and Affordability Applications. 
{\it A presentation given at MOTI/ManTech Meeting, Washington DC,  
October 19, 1996, 11.10am -- 12.10pm.} Copies of transparencies and 
a report distributed to the participants of the meeting (Total 47 pages).\\  

\noindent 
[R6] \  L.J. Kohout. \  \  Tutorial materials prepared for 1996 
Internat. Multidisciplinary 
Conference -- Intelligent Systems: A Semiotic Perspective. National 
Institute of Standards and Technology, Gaithersburg, MD, 20-23 October   
1996. 47 pages. \\

\noindent 
[R7] \  E.~Kim and L.J. Kohout. \  \   Generalized morphisms: 
A tool for evaluation of adequacy of logic
  connectives used in symbolic and soft computing.
 In {\em 1st On-Line Workshop on Soft Computing 
(August 19-30, 1996). Electronic version: 
http://www.bioele.nuee.nagoya-u.ac.jp/wsc1}.
  The Society of Fuzzy Theory and Systems (SOFT), 1996. Also appears in 
hardcopy Proceedings (published by "Nagoya University, Nagoya, 464-01, Japan) 
pp. 222-227. \\

\noindent 
[R8] \  B.~Dubrosky, L.J. Kohout, R.M. Walker, E.~Kim, and 
H.P. Wang. \  \   Use of fuzzy relations for advanced technology 
cost modeling and  affordability decisions.
 In {\it Proc. 35th AIAA Aerospace Sciences Meeting and Exhibit (Reno, 
Nevada, January 6-9,1997)}. AIAA, January 1997, Document AIAA 97-0079, 
pp. 1-12.\\ 

\noindent 
[R9] \  Kohout, L.J. and  Dubrosky, B.  and Wang, H.P. and Zenz, G. and 
Zhang. C. \  \  Decision-making with incomplete information in an 
integrated product 
and process development enterprize -- A management decision tool for cost
modeling and affordability applications. In: {\em Proc. of NSF Grantees 
Conference (Seattle, WA, January 7-10, 19970.} T. Woo, (ed.), pages
401-402. \\ 

\noindent 
[R10] \  E.Kim, L.J.Kohout and B.M.DuBrosky \ \  Linguistic Models of 
Cost-Affordability for Aeronautics Industry based 
on Semiotic Descriptors and Fuzzy Relational Computations. In: {\em
Proc. of 1997 Internat.  Joint Conf. on Information Sciences JCIS'97, March
1-5, 1997, Research Triangle Park, NC}, 
vol. 2 -- Computational Intelligence, Neural Network \& Semiotics. 
P.P. Wang (ed.), pp. 241-244. \\

\noindent
[R11] \   Panel Discussion: {\it Probability, Possibility, and Fuzzy 
Logic Approaches to Affordability in the Aeronautic Industry.} 
Organized by L.J. Kohout. \\ 
Presented at: {\it 1997 Internat.  Joint Conf. on Information  
Sciences JCIS'97}, March 1-5, 1997, Research Triangle Park, NC
Monday, March 3, 14.10pm -- 15.40pm. Co-chaired by: Prof. L.J. Kohout, FSU 
and Dr. Barbara DuBrosky, Pratt \& Whitney. \\ 

\noindent 
[R12] \   Kohout, L.J. and Zenz, G. \  \  Activity Structures and 
Triangle BK-Products of Fuzzy 
Relations -- a Useful Modeling and Computational Tool in Value Analysis 
Studies. In: {\em Proc. of IFSA 1997 -- The world Congress of Internat. 
Fuzzy Systems Association}, vol. IV, pp. 211-216. \\

\noindent
[R13] \   Kohout, L.J. and Kim, E. \  \   Group Transformations of 
Systems of Logic Connectives.
In: {\em Proc. 6th IEEE International Conference on Fuzzy Systems}, 
vol. I, pp.157-162. \\

\noindent
[R14] Zhang, C., Wang, B. and Po-Kang Ting  \  \  Cost Estimation in 
an Integrated Product and Process Development 
Enterprize. Techn. Report No. 97-01 (January 1997). Dept. of Industrial 
Engineering, FAMU-FSU College of Engineering. \\

\noindent
[R15] Noe C.S. \  \  Using BK-Products of Relations for Management 
of Granular Structures 
in Relational Architectures for Knowledge-Based Systems. \\
Ph.D. Dissertation, Dept. of Computer Science, Florida State University, 
May 1997. \\
Major professor: Dr. L.J. Kohout. \\

\noindent
[R16]  \   Kohout, L.J., Kim, E. and DuBrosky, B. \  \  Evaluating 
Affordability of New Technologies by Means of BK-Products
of Fuzzy Relations. \\
An Invited paper presented at AIAA/SAE World Aviation Congress
October 13-16, 1997,  Anaheim, CA.\\
                                          
\noindent
[R17]  \   Kohout, L.J. \  \  Relations and Their Products 
(A Wyllis Bandler Memorial Lecture). 
Invited plenary lecture. 
Presented at: {\it 1997 Internat.  Joint Conf. on Informat.   
Sci. JCIS'97}, March 1-5, 1997, Research Triangle Park, NC\\

\noindent
[R18] Kohout, L.J., Kim, E. \  \  The Role of Semiotic Descriptors 
in Relational Representation of Fuzzy Granular Structures. \\
An Invited paper presented at ISAS '97 Intelligent Systems and
Semiotics: A Learning Perspective (September 22-25 1997) at National 
Institute of Standards and Technology, Gaithersburg.\\

\noindent
[R19] Kohout, L.J., Kim, E. \  \ Semiotic Descriptors in Relational
Computations. \\
Proc. of ISIC/CIRA/ISAS '98 (IEEE International Symposium on 
Intelligent Control (ISIC), International Symposium on Computational 
Intelligence in Robotics and Automation (CIRA), 
Intelligent Systems and Semiotics (ISAS)). National Institute of Standards
and Technology (NIST) Gaithersburg, Maryland U.S.A. (September 14-17,
1998). \\

\noindent
[R20]  Kim, E. and Kohout, L.J. \  \   Generalized morphisms, a new tool for comparative evaluation of
performance of fuzzy implications, t-norms and co-norms in
relational knowledge elicitation. 
{\it Fuzzy Sets and Systems},117(2001), pp. 297-315. \\

\end{document}